\documentclass[fleqn,usenatbib]{mnras}

\usepackage{newtxtext,newtxmath}

\usepackage[T1]{fontenc}

\DeclareRobustCommand{\VAN}[3]{#2}
\let\VANthebibliography\thebibliography
\def\thebibliography{\DeclareRobustCommand{\VAN}[3]{##3}\VANthebibliography}

\usepackage{subfigure} 
\usepackage{xcolor}
\usepackage{graphicx}	
\usepackage{amsmath}	

\newcommand{\kms}{\,km\,s$^{-1}$} 
\newcommand{\kmso}{\,km\,s$^{-1}$\,} 
\newcommand{\mt}[1]{\mathrm{#1}}

\newcommand{\Mo}{$\mathrm{M}_{\odot}$ \,}
\newcommand{\MO}{\ensuremath{\mathrm{M}_{\odot}}}
\def\ud{{\rm d}}

\defcitealias{Hunt2024}{HR24}

\title[Open clusters in the Milky Way]{$N$-body simulations of the open cluster population in the Milky Way and the impact of GMC encounters}

\author[T. G. J\o rgensen et al.]{
Timmi G. J\o rgensen,$^{1}$\thanks{E-mail: timmi.jorgensen@fysik.lu.se}
Ross P. Church,$^{1}$
\\
$^{1}$Lund Observatory, Division of Astrophysics, Department of Physics, Lund University, Box 118, SE-22 100 Lund, Sweden\\
}
\date{Accepted XXX. Received YYY; in original form ZZZ}

\pubyear{2026}

\begin{document}
\label{firstpage}
\pagerange{\pageref{firstpage}--\pageref{lastpage}}
\maketitle

\begin{abstract}
We investigate the population of open clusters in the Solar Neighbourhood and model the effect of encounters with giant molecular clouds (GMCs) over the last 1~Gyr. We combine a Galactic model with $N$-body simulations of 20692 unique clusters in the mass range $[50-24000]$\,\MO. Each cluster is simulated twice: with and without tidal forces from the GMCs. We find that an initial cluster mass function truncated at $\sim 9000$\,\Mo best reproduces the observed mass function evolution. For the age function, the observations show a decline in clusters for ages older than $\sim 1$ Myr, whereas our simulated clusters show a decline after $\sim 50$ Myr. The observed early disruption suggests that some clusters form supervirial, whereas our simulated clusters are created in virial equilibrium. Low-mass ($<600$ \MO) clusters are most sensitive to GMC encounters, which accelerate their disruption in the first $\sim 200$ Myr. After $\sim 500$ Myr, the impact of GMCs becomes irrelevant since the clusters will have been destroyed regardless of whether they experience GMC encounters or not. The survival of intermediate-mass ($600-6000\,\MO$) clusters is significantly reduced by GMCs at ages up to 1\,Gyr. High-mass ($>6000\,\MO$) clusters survive to 1\,Gyr with minimal disruption with or without GMCs. We find that clusters that have had strong GMC encounters within the last $20$ Myr should have tidal tails that are randomly orientated with respect to the Galactic centre.
\end{abstract}




\section{Introduction}
Most stars are believed to form in clustered environments in the form of OB associations or stellar clusters \citep{Lada2003, Krumholz2019}. As a cluster evolves, it continuously loses gas and stars to its surrounding galactic environment. The stars lose mass via stellar evolution \citep{Lamers2010} and two-body relaxation causes stellar evaporation \citep{White1977,Bonnell1998}. However, other than for the rapid mass loss caused by supernova explosions, these internal processes work over longer time-scales than the external tidal forces \citep{Baumgardt2003, Gieles2008}. The evolution of the cluster is therefore mostly dictated by the tidal forces it experiences,  which can be categorised into two types: adiabatic tides and tidal shocks \citep{Renaud2018}. Adiabatic tides occur slowly compared to the cluster's dynamical time-scale, and are produced by the secular variation in the local gravitational potential owing to, for example, an eccentric galactic orbit.  Shocks are produced when the tidal force changes on a short time-scale compared to the cluster's dynamical time-scale, because of rapid interactions with spiral arms, the Galactic disc, and in particular giant molecular clouds (GMCs). 

The dissolution time of clusters impacted by GMCs in the Solar Neighbourhood is a factor of $3.5$ shorter compared to the dissolution from adiabatic tides \citep{Gieles2006}. The presence of GMCs can therefore play an important role when it comes to the disruption and destruction of stellar clusters \citep{Kruijssen2011}. This is especially true for low-mass clusters that are less gravitationally bound and therefore more sensitive to the tidal perturbations produced when GMC encounters occur. To investigate how the cluster population in the Milky Way evolves, it is therefore important to include the effects of GMCs.

The data gathered from the \textit{Gaia} satellite have revolutionised our knowledge of open clusters in the Milky Way \citep{Cantat-Gaudin2022}, and many new open clusters have been discovered and the membership of known clusters has been updated and made more complete \citep{Castro-Ginard2018, Cantat-Gaudin2019, Liu2019, Castro-Ginard2020, Cantat-Gaudin2020, Hunt2021, Castro-Ginard2022, Hunt2023}. The most complete catalogue to date is by \citet{Hunt2024} (hereafter, \citetalias{Hunt2024}) and is based on \textit{Gaia} Data Release 3 \citep{GaiaDR3}, and contains 5647 clusters, where a cluster is defined to be a group of at least 10 stars that are gravitationally bound and more massive than 40 \MO. 

In this paper, we investigate the cluster population of the Milky Way and how it is impacted by GMCs over the last 1 Gyr. We combine a time-varying Milky Way model with $N$-body simulations of 20692 individual stellar clusters in the mass range $[50-24000]$ \Mo and compare our simulated clusters to the recent open cluster catalogue of \citetalias{Hunt2024}. 

The paper is constructed in the following way: Section \ref{sec:MW} describes the Milky Way model and the implementation and distribution of GMCs. In Section \ref{sec:Simulation}, we explain the initial set-up of clusters in the Galactic model and $N$-body simulations, and how we define a cluster. We discuss and compare our cluster population to observations in Section \ref{sec:Results}. We investigate how clusters are impacted by GMCs in Section \ref{sec:GMC_TT} and end with our conclusions in Section \ref{sec:conclusions}.   

\section{Milky Way model}
\label{sec:MW}
We adopt the Milky Way model from \citet{Jorgensen2020}, which consists of a static axisymmetric potential, a Galactic bar, two spiral arms, and individual GMCs that are continuously born and destroyed. 

The axisymmetric potential consists of a thin and thick disc, and a dark matter halo which are all based on models used by \citet{Binney2012}. The density of the spheroid which represents the dark matter halo is given by                                                                                                          
\begin{equation}
\rho_{\rm{DMH}}(r') = \frac{\rho_0}{ \left ( \frac{r'}{r_0}  \right ) ^{\gamma_0} \left ( 1+ \frac{r'}{r_0}  \right )^{\beta_0 - \gamma_0}} \rm{exp} \left [-\left ( \frac{r'}{r_{\rm{cut}}} \right )^2 \right ],
\end{equation}  
where $r' \equiv \sqrt{R^2 + (z/q_0)^2}$. Here, $R$ and $z$ are the cylindrical radius and height, respectively. We adopt Potential I from \citet{Binney2012} with values $\rho_0 = 1.26 \times 10^9$ \Mo kpc$^{-3}$, $q_0 = 0.8$, $\gamma_0 = -2$, $\beta_0 = 2.21$, $r_0 = 1.09$ kpc, and $r_{\rm{cut}} = 1000$ kpc. The density of the thin and thick disc is given by
\begin{equation}
\rho_{\mathrm{disc}}(R,z) = \frac{\Sigma_0}{2 h_z} \mt{exp} \left ( \frac{-|z|}{h_z} \right ) \mt{exp} \left ( -\frac{R}{h_R} \right ) , 
\end{equation}  
where $h_R$ is the radial scale length, $h_z$ is the scale height, and $\Sigma_{0}$ is the disc's central surface density. Both discs have the same scale length of 2.4 kpc, but differ in the scale height where the thin and thick discs have values of 0.36 and 1.0 kpc, respectively.   

\subsection{Galactic bar}
The Galactic bar is represented by a prolate inhomogeneous spheroid \citep{Pichardo2003, Pichardo2012} with a linear density law
\begin{equation}
\rho(a) = \rho_0 (1 - a),
\label{eq:a}
\end{equation}
where $a = (x^2/a_0^2 + y^2/c_0^2 + z^2/c_0^2)^{1/2}$ and $\rho_0$ is the central density. The dimensions of the semimajor and minor axes of the bar are $a_0 = 3.13$ kpc and $c_0 = 1.0$ kpc. Following \citet{Pichardo2012}, the mass of the Galactic bar is $1.6 \times 10^{10}$ \MO.  We adopt a present-day angle between the major axis of the bar and the Sun-Galactic centre line of $-19.2^{\circ}$ \citep{Gaia2023}, and a constant pattern speed of 33.29 \kmso kpc$^{-1}$ \citep{Clarke2022}.    
 
\subsection{Spiral arms}
Our model includes two spiral arms. Each arm is represented by 100 inhomogeneous oblate spheroids with semimajor and minor axes of $1.0$ and $0.5$ kpc. Each oblate spheroid follows the same linear density law as the Galactic bar, i.e. Eq. \ref{eq:a}, with $a_0 = 1.0$ kpc and $c_0 = 0.5$ kpc, and $a = (x^2/a_0^2 + y^2/a_0^2 + z^2/c_0^2)^{1/2}$. The central density of a spheroid, $\rho_0$, is given by
\begin{equation}
\rho_0 = \rho_{02} \, e^{-(R-R_s)/h_R},
\end{equation}
where $R$ is the Galactocentric radius and $R_s$ is the radius where the spiral pattern begins which is equal to the semimajor axis of the Galactic bar, i.e. $R_s = 3.13$ kpc. $\rho_{02}$ is given by \citet{Pichardo2003} as  
\begin{equation}
\rho_{02} = \frac{ 3 M_s}{2 \pi a_0^2 c_0 \sum_{j=1}^{N} e^{-(R_j - R_s)/h_R}},
\label{eq:rho02}
\end{equation} 
where $M_s$ is the total mass of the spiral arms, $R_j$ is the Galactocentric distance of each of the spheroid's centres which is summed over one spiral arm. The total spiral arm mass is $4.0 \times 10^9$ \Mo which is $9.4$ per cent of the total mass in the disc. It should be noted that $h_R$ in Eq. \ref{eq:rho02} refers to the radial scale length of the gas distribution in the Milky Way which has been observed to be $2.0$\,kpc for $R > 3.0$\,kpc \citep{Miville2017}. For our Milky Way model we chose a pitch angle of $14^{\circ}$ based on the findings of \citet{Reid2019}. The present day spiral pattern begins at the same location as the Galactic bar.   

The forces exerted by the prolate and oblate spheroids were pre-calculated numerically to construct force grids as a function of distance and orientation. We used bicubic interpolation in these pre-calculated force grids to estimate the forces from the bar and spiral arms \citep[see][]{Jorgensen2020}. 

Based on open cluster observations in \textit{Gaia} EDR3, \citet{Castro-Ginard2021} found that the pattern speed of the spiral arms in the Milky Way decreases with Galactocentric radius and vary from $\sim 50$ to $\sim 20$ \kmso kpc$^{-1}$. In our model we assume a more simplistic scenario with a constant pattern speed for the spiral arms of 24 \kmso kpc$^{-1}$ which is in good agreement with the average pattern speed of the three nearest spiral arms estimated by \citet{Castro-Ginard2021}.      

The total mass and density profile of our Galactic model are in agreement with estimates of \citet{McMillan2017}, \citet{Eadie2019} and \citet{Cautun2020}. Our rotation curve has a circular velocity of $v_\phi=223.5\pm 8.3$\,\kms\ in the Solar Neighbourhood, which corresponds well with the estimate by \citet{Fedorov2025} of $229.63\pm 0.30$\,\kms.  The circular velocity is a proxy for the bulk local potential; since our sample is relatively local it is important to reproduce it well.

Our Galactic model with a static bar and spiral arms is relatively simple, and could potentially be improved by adopting other prescriptions for these structures such as an evolving bar and spiral pattern \citep[see][]{Hunt2018,Hunt2019}. However, the evolution of the structures of the Milky Way over the last 1 Gyr is not well known.  Our approach here is to investigate the properties of a reasonable model of the local Galactic environment on open clusters, rather than to carry out a parameter study to investigate the sensitivity of the model to its assumptions.
 
\subsection{GMCs}
To create a realistic model of the stellar cluster lifecycle in the Milky Way, it is important to implement a GMC population that matches observations in lifetime, mass, size, and spatial distribution. Here we list the implementation and representation of GMCs in our Milky Way model. 

\subsubsection{Evolution}
Based on simulations of a Milky Way galaxy, \citet{Benincasa2020} found GMC ages of $5-7$ Myr, with less than 2 per cent of the GMC population having lifetimes longer than 20 Myr. The simulations of \citet{Ni2025} found a larger spread, from a few to several tens of Myr. This spread is largely dependent on the environment of the GMCs, the star formation efficiency, and stellar feedback. \citet{Ni2025} also found that there is a significant correlation between the maximum mass of a GMC and its lifetime for masses in the range of $\sim 10^5 - 10^7$ \MO. Based on the age dispersion of stars in stellar clusters, the lifetimes of molecular cloud complexes are believed to be of the order of their dynamical time-scales \citep{Elmegreen2000}. Furthermore, the lack of $\sim 5 - 10$ Myr old stars in star-forming regions in the Solar Neighbourhood indicates that GMCs form stars in a short period of time before dispersing \citep{Hartmann2001}.  
Assuming that GMCs are disrupted by feedback from massive stars, \citet{Murray2011} estimated lifetimes of $27 \pm 12$ Myr which are supported by theoretical studies \citep{Williams1997,Matzner2002} which predict GMC lifetimes in the range of $10-30$ Myr. 

The typical estimated age of GMCs can vary substantially and is a combination of several factors that is likely unique for each GMC. In our model, we assume that all GMCs have the same lifetimes which we set to be 20 Myr which we believe to be a good median age in terms of what observations and theoretical studies suggest. Adopting a similar prescription as \citet{Gustafsson2016}, each GMC follows a parabolic mass evolution over its lifetime of the form
\begin{equation}
M(t) =  \left [ - \left ( \frac{t-t_0}{10 \, \mt{Myr}}  \right )^2 +  \frac{t-t_0}{5 \, \mt{Myr}} \right ] \cdot M_0,
\label{eq:M_evolution}
\end{equation}
where $M_0$ is the maximum mass of the GMC, $t$ is the current time, and $t_0$ is the birth time of the GMC. As gas cools in the Galaxy, GMCs continuously form and disperse as a consequence of stellar feedback, and is the reasoning behind the parabolic mass evolution. The ages of GMCs in our model follow a flat distribution where the GMCs are continuously born and destroyed, following the evolution of Eq.~\ref{eq:M_evolution}. As such, the total number of GMCs present in the model is constant with time. A \citet{Plummer1911} sphere is used to represent each GMC with a virial radius, $R_v$, which is related to the current mass of the GMC. The mass-size relation for our GMCs is based on simulations by \citet{Hopkins2012}, who suggest $R_v \propto M^{1/2}$, which is also supported by observations \citep{Larson1981,Lada2020}.  We normalise this relationship using the catalogue of Milky Way GMCs of \citet{Miville2017}, who used $^{12}$CO observations obtained over a period of 20 years by \citet{Dame2001}. \citet{Miville2017} applied a hierarchical cluster identification method to a Gaussian decomposition of the CO map in order to produce their catalogue. We fit their estimated masses and virial radii of GMCs with masses $\geq 10^5$ \Mo to estimate a mass-size relation given by   
\begin{equation}
R_v(t) =  23.9 \left( \frac{M(t)}{10^5 \, \mathrm{M}_{\odot}} \right )^{1/2} \, \mt{pc}.
\end{equation}  

\begin{figure}
 \includegraphics[width=\columnwidth]{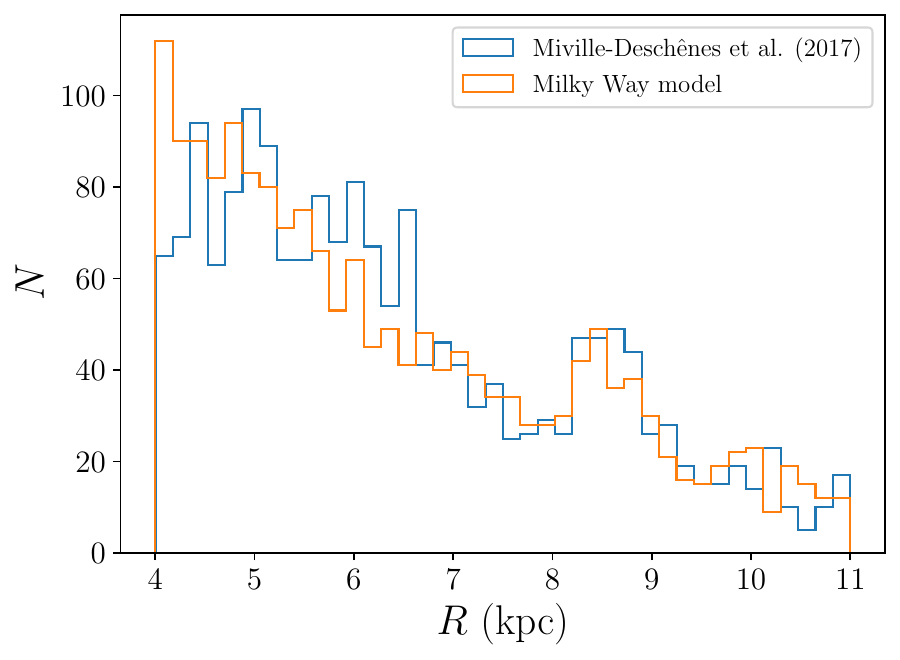}
 \caption{The radial distribution of GMCs with masses $\geq 10^5$\,\Mo (blue) from the catalogue of \citet{Miville2017}, compared to the distribution in our Milky Way model (orange).}
\label{fig:N_GMC}
\end{figure}

\subsubsection{Spatial distribution}
One would expect the radial spatial distribution of the GMCs to follow that of the gas disc in the Milky Way. This is indeed also the case, however, there seems to be a distinct difference in the number of GMCs distributed in the inner and outer parts of the Milky Way. The inner and outer regions are separated by the solar circle. A substantial fraction of molecular gas in the Milky Way is believed to be located in a ring in the Galactocentric range $3 \lesssim R \lesssim 7$ kpc \citep{Bronfman1988, Nakanishi2006}. This ring is likely what creates this distinct difference between the number of GMCs in the inner and outer parts of the Milky Way. Fig. \ref{fig:N_GMC} shows the number distribution as a function of Galactocentric radius for GMCs with masses greater than $10^5$ \Mo in the range of 4 to 11 kpc. The inner and outer parts of the Milky Way contain 1405 and 393 GMCs, respectively. We are only interested in this Galactic range, since we want to compare our results with the cluster catalogue of \citetalias{Hunt2024} which is only complete for distances below $\sim 3$ kpc. For our Galactic model, the spatial distribution of GMCs is split into two separate distributions which are separated at the solar circle, for which we adopt a value of $R = 8.178$ kpc \citep{GRAVITY2019}. Both distributions follow the same radial scale length of the gas of $2.0$ kpc. The model follows the observations quite well. However, there is an overestimate of the number of GMCs in the range $4 - 5$ kpc and a slight underestimate of GMCs around the solar circle.    

In our Milky Way model, the GMCs are born in the Galactic disc within 50 pc of the locus of the spiral arm pattern. This distribution is based on \citet{Gustafsson2016}, who tested whether a random distribution of GMCs in the disc had an impact on the statistical properties of cluster orbits compared to GMCs being born along the spiral arms. They found that there was less than $2$ \kmso difference in the velocity dispersion; therefore we do not think that a different choice of distribution around the spiral locus would significantly change our results. The exact location of the GMCs of the Milky Way is not known, due to the fact that we live in the Galactic disc and therefore cannot get a face-on view of our Galaxy; however, observations of GMCs in other galaxies, \citep[e.g. M51][]{Colombo2014} have shown that a fraction of GMCs can be located between the spiral arms. Because of this, we allow for 10 per cent of the GMCs to be born randomly in the disc, while still following the radial distribution of the gas.  
Each GMC is born with an initial Gaussian velocity dispersion of 7 \kmso in each of the three spatial directions, similar to the Milky Way model of \citep{Jorgensen2020}, which showed good agreement with the observations of \citet{Holmberg2009}.

\subsubsection{GMC mass distribution}
\citet{Rice2016} found that the mass distributions of the GMCs in the inner and outer parts of the Milky Way are not identical. The inner part of the Milky Way not only contains more GMCs, but they are generally also more massive. \citet{Rice2016} found that the mass function of the GMCs in the inner part of the Milky Way is best described by a truncated power-law, whereas the outer part is best represented by a non-truncated power-law. The GMC catalogue of \citet{Rice2016} only contains $\sim 25$ per cent of the total molecular mass, whereas the catalogue of \citet{Miville2017} is almost complete with 98 per cent of the total $^{12}$CO emission observed. We therefore used the catalogue of \citet{Miville2017} to investigate if the difference in the mass function of the GMCs in the two regions is also present in these data. 
The cumulative mass function of a power-law scales as $\propto M^{\beta +1}$, whereas the cumulative mass function of a truncated power-law \citep{Williams1997} is given by
\begin{equation}
N(M' > M) = N_0 \left [  \left (  \frac{M}{M_{\rm{max}}} \right )^{\beta + 1} -  1 \right ],
\label{eq:GMC_PL}
\end{equation}
where $\beta$ is the power-law slope of the differential mass function, $M_{\rm{max}}$ is the maximum mass of the distribution, and $N_0$ is the number of objects more massive than $2^{1/(\beta+1)} M_{\rm{max}}$, which is the mass where the distribution deviates from a power-law. From the data catalogue of \citet{Miville2017} we divided the 1798 GMCs with masses $\geq 10^5$ \Mo into two groups: an inner and outer distribution which are separated by the solar circle. Each group was then fitted with a cumulative distribution given by Eq. \ref{eq:GMC_PL} where the GMCs were binned into equal sized bins of 60 and 20 GMCs for the inner and outer groups, respectively. The initial error on each bin is given as a statistical error by $\sigma_{\rm{bin}} = \sqrt{\frac{n_{\rm{bin}} (n_{\rm{tot}} -n_{\rm{bin}})}{{n_{\rm{tot}}}}}$, where $n_{\rm{bin}}$ is the number of GMCs in the bin and $n_{\rm{tot}}$ is the total number of GMCs in the population. After assigning these errors, each bin was then scaled due to the error propagation of representing the data as a cumulative distribution. The error estimate of the mass of each bin was based on the standard deviation of the distribution in each bin. 

\begin{table}
 \begin{tabular*}{\columnwidth}{@{\extracolsep{\fill}} lcccccccc}
  Region & $\beta$ & $M_{\mathrm{max}}$ & $N_0$ & $N_{\rm{GMC}}$ \\
           &         & $(10^6$ \MO) &    \\
  \hline
  Inner(lower) & $-1.45\pm 0.02$ & $2.56\pm 0.12$ & $440\pm 45$ & $1374$ \\   
  Inner(upper) & $-3.70\pm 0.09$ & $-$            & $-$         & $31$   \\
  Outer(lower) & $-1.66\pm 0.02$ & $3.3\pm 0.4$   & $43\pm 7$   & $387$  \\
  Outer(upper) & $-1.07\pm 0.10$ & $6.8\pm 0.9$   & $129\pm 60$ & $6$    \\
  \hline
 \end{tabular*}
 \caption{The best fit parameters for the cumulative mass distribution of the GMCs with masses $\geq 10^5$ \Mo in the catalogue of \citet{Miville2017}. Three out of the four distributions are best described by a truncated power-law given by Eq. \ref{eq:GMC_PL}. The inner(upper) distribution is best described by a non-truncated power-law, which is indicated by the omission of the values $M_{\mathrm{max}}$ and $N_0$. $N_{\rm{GMC}}$ indicates the number of observed GMCs in each region.} 
 \label{table:GMC_fits}
\end{table}

\begin{figure}
 \includegraphics[width=1.0\columnwidth]{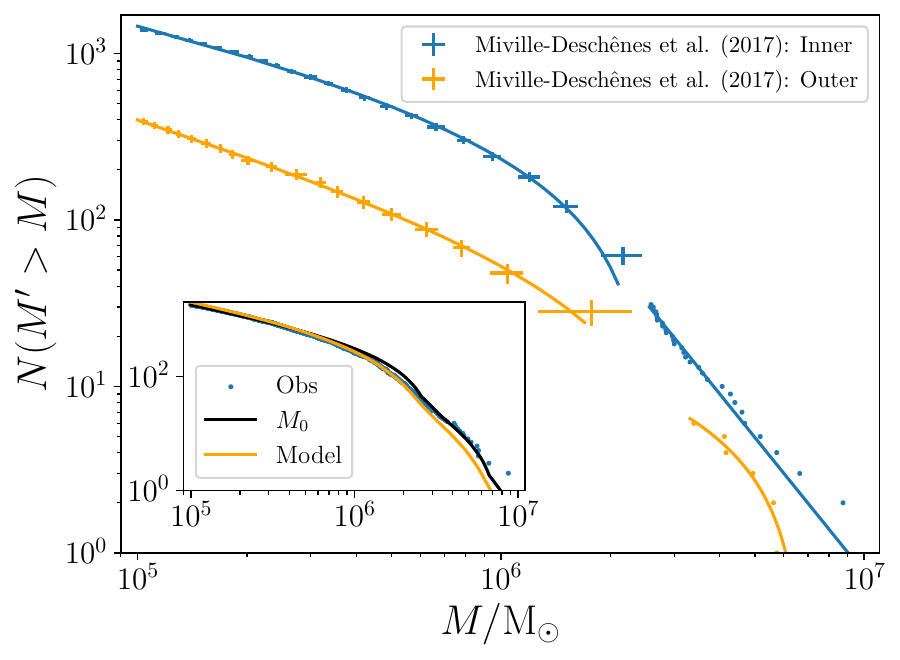}
 \caption{Fits to the cumulative mass distribution of GMCs from \citet{Miville2017}.  Blue points show data from inside the solar circle ($R_{\rm gal}<8.178\,{\rm kpc}$); orange points from outside the solar circle. The lower mass regions are binned, whereas the upper mass regions are un-binned due to the low number of GMCs. A truncated power-law is the best description for all populations, except the inner upper region, which is best described by a non-truncated power-law. The subplot shows the cumulative distribution of GMCs from \citet{Miville2017} and our Milky Way model where both the $M(t)$ and $M_0$ distributions from Eq. \ref{eq:M_evolution} can be seen and are based on the parameters of the fits which are given in Table \ref{table:GMC_fits}.}
\label{fig:Fit_GMC}
\end{figure}

By fitting the cumulative GMC mass function to a truncated power-law, any GMCs with masses greater than the estimated truncation mass, i.e. $M_{\rm{max}}$, will not be represented by the function. In other words, it will not be possible for us to draw GMC masses from the distribution given by Eq. \ref{eq:GMC_PL} that are greater than $M_{\rm{max}}$ even though they exist in the data. In order to make this possible, we divided each distribution function into two separate fits: one for GMCs below the truncation mass, which we refer to as the lower region, and one for the GMCs masses above which we refer to as the upper region. To separate each region, we made an initial estimate of the truncation mass for the GMC population in both the inner and outer parts of the Galaxy. From here, we only re-binned the GMCs which had masses below the estimated truncation mass, i.e. the lower mass region. The GMCs with masses above the truncation mass were classified as part of the upper mass region and were not binned due to the small number of GMCs. The fitting of the inner and outer population can be seen in Fig. \ref{fig:Fit_GMC} and the resulting estimated parameters are listed in Table~\ref{table:GMC_fits}. Both of the lower mass regions of the GMC distribution are best described by a truncated power-law, but with significantly different slopes. The inner part of the Galaxy has a slope of $-1.45\pm 0.02$, whereas the outer part is steeper with a value of $-1.66\pm 0.02$. The lower mass regions contain the majority of the GMCs and we see that there is a larger fraction of high-mass GMCs in the inner part of the Galaxy which is consistent with the results of \citet{Rice2016}. The increased molecular gas in the Galactic rings is likely what allows the formation of higher-mass GMCs in this part of the Galaxy. For the upper mass regions of the inner and outer part of the Galaxy, we find that the inner part is best described by a non-truncated power-law with a slope of $-3.70\pm 0.09$ which reveals that the most massive GMCs with masses around $10^7$ \Mo are very rare objects. For the outer part of the Galaxy, a truncated power-law best describe the GMC distribution, with a slope of $-1.07\pm 0.10$ and a truncation mass of $(6.8\pm 0.9) \times 10^6$\,\MO. 

\begin{table}
 \caption{The parameters of the Milky Way model components. The mass listed for the dark matter halo is the mass contained within a spherical radius of 20 kpc.}
 \label{table:MW_model}
 \begin{tabular*}{0.50\textwidth}{@{\extracolsep{\fill}} ccc}
  \hline
  \hline
  & Milky Way model & \\
  \hline
 \end{tabular*}
 \begin{tabular*}{0.50\textwidth}{@{\extracolsep{\fill}} lccc}
    & Mass & $h_R$ & $h_z$ \\
    & $(\rm{M}_{\odot})$ & $(\rm{kpc})$ & $(\rm{kpc})$ \\	
  \hline
  \\ 
  Dark matter halo & $1.14 \times 10^{11}$ & -  & - \\
  Thin disc & $4.1\times 10^8$ & 2.4  & 0.36 \\
  Thick disc & $3.7\times 10^{10}$ & 2.4  & 1.0 \\
  Bar & $1.6 \times 10^{10}$ & -  & - \\
  Spiral arms & $4.0 \times10^{9}$ & 2.0  & - \\
  GMCs & $1.21 \times 10^9$ & 2.0  & - \\	
  \hline
  \end{tabular*}
\end{table}

From these fits, we recreated the GMC mass distribution for the inner and outer parts of the Milky Way. Because our individual GMCs have masses that evolve with time (Eq.~\ref{eq:M_evolution}), we need 2697 GMCs in our Galactic model in order to match the 1798 observed GMCs with masses $\geq 10^5\,$\Mo from \citet{Miville2017}. Here we assume that the observed GMCs are all at their peak mass, i.e. $M_0$ from Eq. \ref{eq:M_evolution}. This means that our modelled GMCs will always have masses that are somewhat below the observed values. The combined cumulative GMC mass distribution can be seen as the subplot in Fig. \ref{fig:Fit_GMC}, where the observed, $M_0$, and modelled GMCs are shown. The 2697 GMCs in our model produces a total mass of $1.21 \times 10^9$ \Mo which agrees well with the observed estimates of molecular gas mass of the Milky Way \citep{Rice2016, Miville2017}. 

Finally, the GMCs feel the forces of other GMCs, and in that sense they are therefore not treated as test particles, unlike the clusters in the Milky Way model. All the parameters and components of our Milky Way model are listed in Table~\ref{table:MW_model}.           

\section{Simulations}
\label{sec:Simulation}
We wish to model the evolution of the Milky Way stellar cluster population over the last 1 Gyr. The evolution of the population is dictated by the Galactic environment in which the clusters reside. This is especially the case for low-mass clusters, which are more sensitive to the strength of and variations in their local gravitational tidal field. We are mainly interested in the gravitational perturbations caused by nearby GMCs, which can contribute to the dissolution and destruction of stellar clusters. To investigate the effects of GMCs in particular, we follow the same approach that we took for M51 in \citet{Jorgensen2025} and run two $N$-body simulations for each unique cluster: one where the tidal effects of the GMCs are present, and one where this effect is ignored but the {\it orbits} of the clusters still follow the full potential including the effects of the GMCs. We can thus make a comparison between the two populations, but also a direct comparison of the tidal effects from the GMCs on each individual cluster. For the rest of this paper, we refer to clusters that feel the tidal effects of GMCs as C$_{\rm{R}}$ ({\it realistic}) clusters, and clusters that do not experience tidal effects from GMCs as C$_{\rm{N}}$ ({\it no GMC}) clusters.

\subsection{Galactic simulations}
The simulations of clusters in our Milky Way model start out 1\,Gyr in the past and continue up until the present time. In the model, clusters are born at a constant rate, which reflects the constant star formation rate in the Milky Way during this time period \citep{Haywood2016}. Clusters are represented as test particles and form with the same spatial and velocity distributions as GMCs. We integrate their orbits in the Galactic model with a 4th order Runge-Kutta integrator and a time step of 0.1 Myr. We ran 36 simulations, each with a unique random seed, and with 10\,000 clusters in each simulation. At the end of the simulations, we extracted the properties of the clusters within 3\,kpc of the solar location, which in our model is $x = 8.178$ kpc, $y = 0.0$, and $z = 20.8$ pc \citep{Bennett2019}. We find a total of 20692 clusters within 3\,kpc, which corresponds to $\sim 6$ per cent of the total cluster population that we followed in the Milky Way model.

\subsection{$N$-body simulations}
To perform the $N$-body simulations of the clusters, we used the code \textsc{nbody6tt} \citep{Renaud2011}, which is a modified version of \textsc{nbody6} \citep{Aarseth2003}. \textsc{nbody6tt} allows the use of a tidal tensor which can describe the local galactic environment of a stellar cluster. The tidal tensor is defined as
\begin{equation}
\boldsymbol{T}_{ij} = -\frac{\partial^2 \Phi}{\partial x_i \partial x_j},
\label{eq:tt}
\end{equation}
where $\Phi$ is the local galactic gravitational potential. We constructed two tidal tensors for each cluster in the Milky Way model: one realistic tidal tensor (C$_{\rm{R}}$) and a tidal tensor where the tidal contribution from the GMCs had been removed (C$_{\rm{N}}$). The tidal tensor was constructed by numerically differentiating the force acting on the cluster for each time step within a cube of edge length $2\delta$. We chose $\delta=15\,$pc which is approximately a few half-mass radii for most clusters in our simulations, hence we measure the tidal force on the relevant length scale for the cluster. All structures that contribute to the force are included in this calculation, and most of them are extended objects that vary on a much larger length scale and hence whose effects can be accurately described by the tidal tensor. However, when clusters experience close GMC encounters, the linearised tidal force is not a sufficient approximation to the external force on the stars. In such cases, we directly simulated the GMC encounters in \textsc{nbody6tt} which is further discussed in Section \ref{sec:GMC_sim}. 

\begin{table}
 \begin{tabular*}{\columnwidth}{@{\extracolsep{\fill}} lcccc}
  $N_{\rm{star}}$ & Mass & $N_c$  &  $n_{\rm factor}$ \\  
   & (\MO) & \\
  \hline
$100 - 1000$    & $\sim 50 - 600$     & $13227$ & $10.12$ \\ 
$1000 - 10000$  & $\sim 600 - 6000$   &  $6350$ & $2.11$  \\ 
$10000 - 40000$ & $\sim 6000 - 24000$ &  $1115$ & $1$ \\ 
    \hline
 \end{tabular*}
 \caption{The number of clusters, $N_c$, simulated within each mass or equivalently star-number, $N_{\rm{star}}$, interval. To investigate the combined cluster population at the end of the simulations, a normalisation is performed where $n_{\rm factor}$ is the normalisation factor which indicates the number of clusters each of our simulated clusters represents from the ICMF population.}
 \label{table:cluster_range}
\end{table}

\subsection{Cluster setup}
\label{sec:cluster_setup}
In order to investigate the stellar cluster population in the Milky Way, we need our simulated clusters to span a wide range in initial masses. We wish to compare our clusters to the catalogue of \citetalias{Hunt2024}, in which cluster masses range from 40 to $2.3 \times 10^4$\,\MO. This roughly corresponds to clusters which have between $10^2$ and $4 \times 10^4$ stars. The initial cluster mass function (ICMF) can be described by a power-law index of $-2$ \citep{Krumholz2019}, which means that the majority of clusters will be born as low-mass clusters. We wish to model a sufficiently large number of higher-mass clusters to resolve the mass function, but avoid the computational cost of sampling too many lower-mass clusters. To that end, we divide the clusters into mass ranges which were rescaled to recreate our chosen ICMF when we compared the population to the observations. This allows us to cover the entire cluster mass range while simultaneously being able to investigate a sizeable sample of high-mass clusters. The mass ranges are listed in Table \ref{table:cluster_range}; within each range the masses are drawn from a power-law distribution with slope $-2$. The approach also allows us to modify the power-law ICMF by e.g.~introducing a high-mass cut-off.

\begin{figure}
\centering
\includegraphics[scale=0.5]{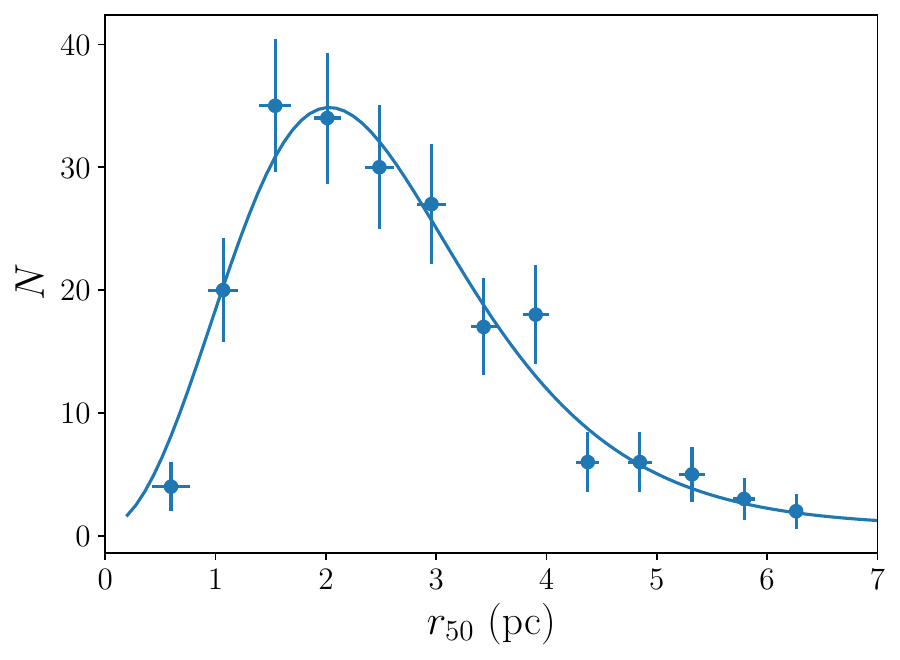}
\caption{Determination of the ICRF.  Points show the distribution of $r_{50}$ distribution for high-quality clusters younger than $10$ Myr from \citetalias{Hunt2024}, in bins of width 0.5\,pc.  Our best-fit distribution (solid line) is given by Eq. \ref{eq:ICRF}.}
\label{fig:ICRF_fit}
\end{figure}

Newborn stellar clusters do not only vary in mass, but also in size. To estimate the initial radius distribution, we used the high-quality clusters from \citetalias{Hunt2024}, since these have a more reliable parameter estimation. Our simulations start after gas has been expelled from the clusters and the embedded phase has finished.  Since the dynamical time-scales of the clusters after gas expulsion are of order Myr, the \citetalias{Hunt2024} clusters with ages less than $10$\,Myr are equivalent to dynamically young clusters in our simulations, and hence a good representation of the initial cluster size distribution. Out of the 3530 high-quality clusters there are 209 younger than 10\,Myr. \citetalias{Hunt2024} do not estimate half-mass radii, but rather measure the radii within which 50 per cent of the clusters' stars are contained, $r_{50}$. We neglect mass segregation and take $r_{50}$ to be equal to the initial half-mass radii.  The $r_{50}$ distribution of the 209 clusters can be seen in Fig. \ref{fig:ICRF_fit} from which we fit an initial cluster radius function (ICRF) given by
\begin{equation}
\rm{ICRF}(r) = \frac{r^2 \xi_1}{e^{r/\xi_0} + \xi_3} + \xi_2,  \quad  r \in [0.4,6.0] \,\, \rm{pc},
\label{eq:ICRF}
\end{equation}   
where $\xi_0 = 0.7$\,pc, $\xi_1 = 217.2\,{\rm pc^{-2}}$, $\xi_2 = 0.76$, and $\xi_3 = 8.1$. The ICRF is defined for clusters with initial radii between 0.4 to 6.0\,pc; however, \citetalias{Hunt2024}'s high-quality sample does not contain any clusters with masses above 600\,\Mo and radii below 1.0\,pc. Because of this, for clusters with initial masses below 600\,\Mo we draw a radius from the whole range of the ICRF, whereas for cluster masses above 600\,\Mo we restrict the radii to the range of 1.0 to 6.0\,pc.  Other than this difference in size for low-mass clusters, we only see a very weak correlation between cluster mass and $r_{50}$ in the \citetalias{Hunt2024} young cluster sample: hence we do not implement any other dependence between the initial cluster radii and mass in our models.

In our $N$-body simulations, the stellar masses are initial distributed following the \citet{Kroupa2001} mass function in the range $0.1$ to $50$ \MO. All stars are born as single stars, with stellar evolution enabled using the prescription of \citet{Hurley2000, Hurley2002} with a metallicity of $Z=0.02$. The initial spatial arrangements of the stars follow a \citet{Plummer1911} distribution. Our model clusters are born in virial equilibrium; i.e. they have a virial ratio
\begin{equation}
Q = \frac{T}{|W|}=\frac{1}{2},
\end{equation}
where $T$ and $W$ are the total kinetic and gravitational potential energies of the cluster.

\subsection{How to define a cluster}
\label{sec:HuntMethod}
The internal evolution of a stellar cluster is driven by two-body relaxation. High-mass stars sink towards the cluster centre, and low-mass stars will diffuse outward to the outer parts of the cluster. Along with tidal heating, this process causes the cluster to fill its tidal radius.  Stars that reach the tidal radius with positive energies may be able to escape the gravitational potential of the cluster. The tidal radius, $r_t$, can be defined by the maximum eigenvalue, $\lambda_{\rm{max}}$, of the tidal tensor \citep{Renaud2011} according to
\begin{equation}
r_t = \left ( \frac{G M_{\rm c}}{\lambda_{\rm{max}}} \right )^{1/3}.
\label{eq:rt}
\end{equation}
Here $M_{\rm c}$ is the cluster's mass and $G$ is the gravitational constant. We consider a star to have escaped the cluster when its total energy is positive and it is located at a radius greater than $r_t$ from the cluster's centre.

When using Eq. \ref{eq:rt} to estimate a cluster's tidal radius, it is possible that all the eigenvalues are temporarily negative, i.e. the cluster is only experiencing compressive tides. In such cases, we evaluate the tidal radius at the last point in time for which the tidal tensor has a positive eigenvalue. In order to make the best comparison between the C$_{\rm{R}}$ and C$_{\rm{N}}$ clusters, we used the tidal tensor of the C$_{\rm{N}}$ clusters to evaluate $\lambda_{\rm{max}}$ for both $N$-body cluster versions. We define a cluster to be destroyed if it contains less than 10 stars by the end of its simulation.

Defining the cluster by its tidal radius and stellar energies is straightforward for simulated clusters. However, it is much more difficult to determine which stars are gravitationally bound to a cluster from observations, since the velocity uncertainties are comparable in magnitude to the internal motions.  Furthermore, when comparing our model clusters to observations it is important to use a self-consistent means of determining cluster membership.  We therefore implement the membership definition of \citetalias{Hunt2024} (henceforth the {\it Hunt method}, summarised below) and investigate how it differs from our energy-based definition.

\citetalias{Hunt2024} estimate the enclosed mass of each cluster as a function of radius, $M_{\rm{obs}}(r)$. By inverting the equation for the tidal radius, they calculate the theoretical Jacobi mass, $M_J(r)$, the mass that a cluster would be required to have as a function of radius $r$. By comparing $M_{\rm{obs}}(r)$ and $M_J(r)$, they estimate $r_t$ as the radius for which the enclosed mass $M_{\rm{obs}}(r_t) = M_J(r_t)$.  The cluster mass is then taken to be $M_{\rm{obs}}(r_t)$ since the stars outside this radius will be unbound. If a cluster has $M_{\rm{obs}}(r) < M_J(r)$ for all radii, then the cluster is not bound and is classified as a moving group. If $M_{\rm{obs}}(r) > M_J(r)$ for all radii, then all stars are bound to the cluster and the tidal radius is calculated using the total cluster mass.

\begin{figure}
  \includegraphics[width=\columnwidth]{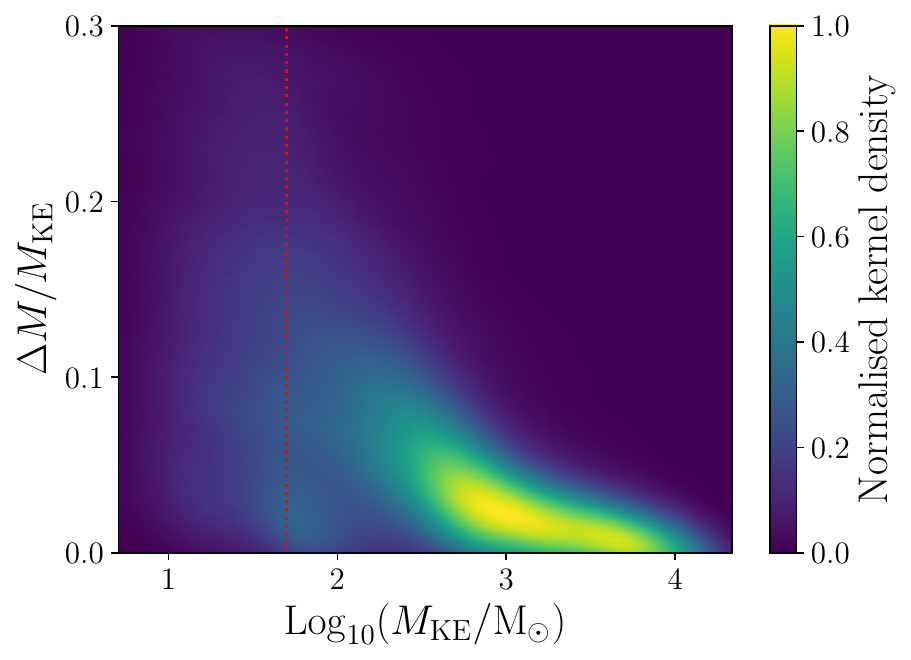}
 \caption{The fractional difference in mass, $\Delta M/M_{\rm KE}$, between the kinetic energy method and the Hunt method for determining the mass of a stellar cluster, plotted as a function of the mass estimated using the kinetic energy method, $M_{\rm KE}$. The plot has been smoothed using a kernel density estimator which have been normalised to a maximum value of 1. The red dotted line shows the lowest mass in our initial cluster mass function.}
\label{fig:dM_kernel}
\end{figure}

By using the Hunt method, the bound cluster members can easily be classified without needing to rely on kinematics. However, stars that are located beyond the tidal radius but are still bound to the cluster, i.e. have negative energies, are no longer considered part of the cluster. Cluster masses estimated using the Hunt method will therefore be lower than if the energies of the stars were also considered. To investigate the impact of the derived cluster mass, we evaluated our simulated clusters using both methods. 
We calculated the fractional mass change $\Delta M/M_{\rm{KE}}= (M_{\rm{KE}} - M_{\rm{H}})/M_{\rm{KE}}$, where $M_{\rm{KE}}$ is the mass calculated considering the stellar energies and $M_{\rm{H}}$ is the cluster mass calculated by the Hunt method. 
Fig. \ref{fig:dM_kernel} shows $\Delta M/M_{\rm{KE}}$ as a function of $M_{\rm{KE}}$, where a normalised kernel density\footnote{We use {\tt scipy.stats.gaussian{\_}kde} with default parameters.} has been used to smooth the distribution. The vertical dotted line represents the lowest cluster mass of the ICMF of 50\,\MO. We see that the clusters which have the largest difference in mass are the low-mass clusters, however, the majority of these are contained to a difference of 20 per cent or less. In terms of morphology, the clusters evaluated by the Hunt method will be more spherical, due to the hard cut off in radius which is set by the tidal radius, but this will mostly affect the low-mass stars since they are more common in the outer parts of the clusters. The majority of clusters have mass differences less than 10 per cent, which is less than the stated uncertainties in the \citetalias{Hunt2024} catalogue. 

\subsection{Simulating close GMC encounters}
\label{sec:GMC_sim}
The linear tidal tensor can be a poor representation of the tidal force if a cluster is experiencing a sufficiently close GMC encounter. This happens when the separation between the cluster and GMC is small, and becomes significantly worse the closer the two objects are to each other. By comparing the approximation of the tidal tensor to the true tidal force, we found that for all GMC encounters, there is a less than 20 per cent force error when the separation at closest approach is greater than $60$ pc. These GMC encounters, if strong enough to do significant work on the clusters, are therefore poorly represented in the tidal tensor. A strong GMC encounter can be defined using the approximation for the fractional change in the energy of a cluster, given by \citet{Gustafsson2016} as 
\begin{equation}
\delta_E = \frac{8 G M_{\rm{GMC}}^2 r_{\rm h}^3}{3 M_{\rm c} b^4 V^2}.
\label{eq:dE}
\end{equation}
Here, $M_{\rm{GMC}}$ is the mass of the GMC, $r_{\rm h}$ is the cluster half-mass radius, $b$ is the impact parameter, and $V$ is the relative velocity between the two objects at infinity. GMC encounters which generate $\delta_E \lesssim 0.01$ have little to no effect on the cluster's evolution \citep{Gustafsson2016}, and we therefore define encounters as strong when the separation at closest approach is below 60 pc and $\delta_E \ge 0.01$. GMC encounters with large $\delta_E$ can take place at larger distances, but these are well approximated in the tidal tensor. Similarly, GMC encounters with $\delta_E < 0.01$ are included in the tidal tensor, but have little effect on the cluster.  

Strong encounters need to be followed directly in \textsc{nbody6tt}. When a cluster  experiences a strong GMC encounter, we replace the tidal tensor with the corresponding tensor used for the C$_{\rm{N}}$ simulation. By doing so, we are effectively removing any GMC contribution to the tidal tensor and the GMC encounter is instead followed directly in \textsc{nbody6tt}. During a strong encounter, the tidal field of the cluster will be dominated by the effects of the strong GMC encounter. The contribution of the other GMCs to the tidal field is small compared to the strong encounter and can therefore be ignored. Because our Milky Way model and \textsc{nbody6tt} do not share the same reference frame, we had to estimate the initial position and velocity for the GMC as it is born in \textsc{nbody6tt}, which we did by integrating the orbit backwards from the point of closest approach using a 4th order Runge-Kutta integrator. At the start of each $N$-body simulations, the GMCs that will produce a strong encounter are loaded into \textsc{nbody6tt} with information of their initial position, velocity, maximum mass, and time of birth. When the $N$-body simulation reaches the birth time of each GMC, they become active and evolve in mass according to Eq. \ref{eq:M_evolution} until they reach the end of their life after 20 Myr.     

\begin{figure}
 \includegraphics[width=\columnwidth]{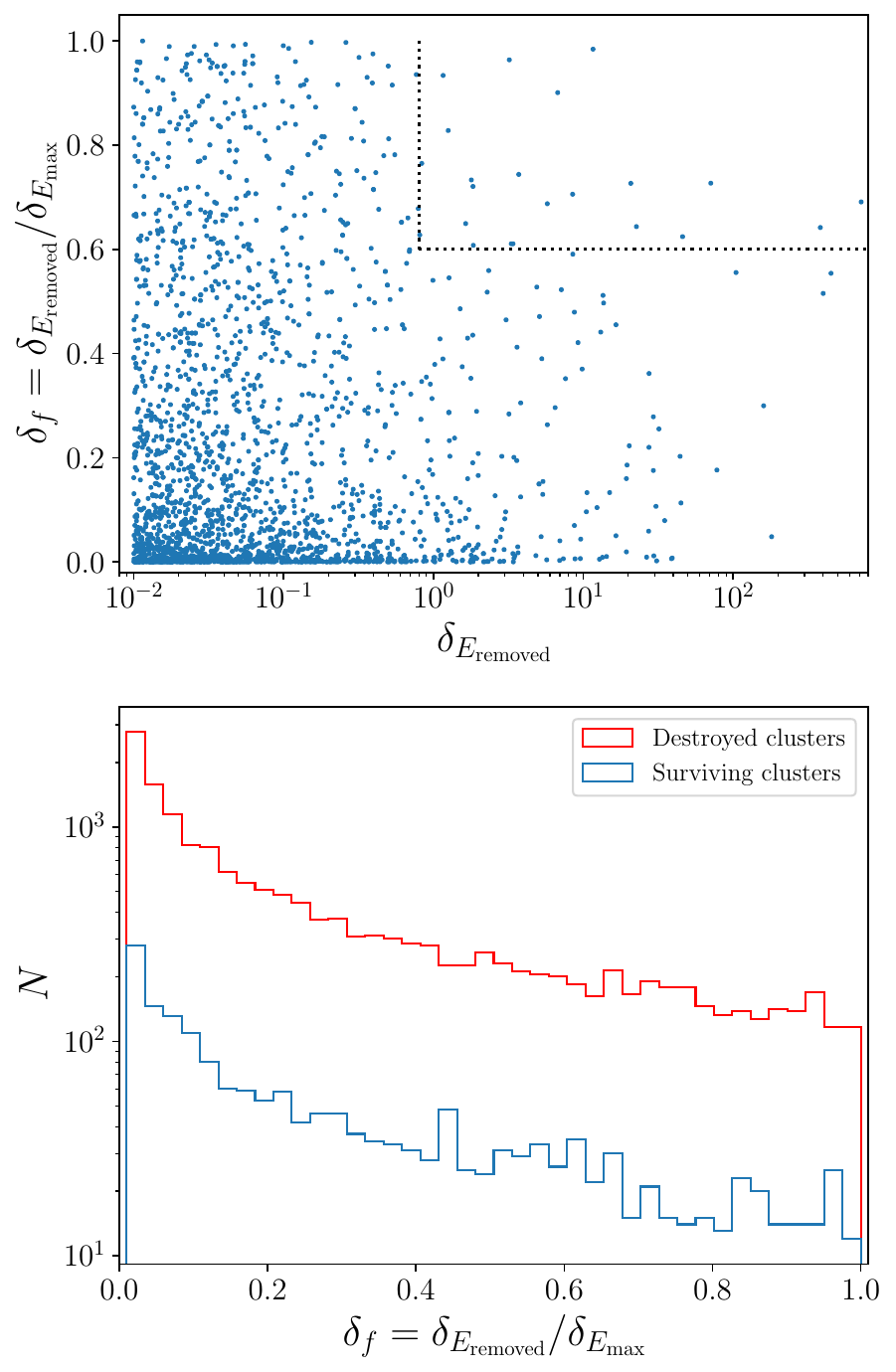}
 \caption{\textit{Top:} The energy ratio, $\delta_f$, of the strongest GMC encounter $\delta_{E_{\rm{max}}}$ and the removed GMC encounter $\delta_{E_{\rm{Removed}}}$ for clusters which have several GMC encounters in the same period of time, as a function of $\delta_{E_{\rm{Removed}}}$. Here, only the surviving clusters are shown. \textit{Bottom:} The distribution of $\delta_f$ for all 7534 unique clusters which have had one or more GMC encounters removed in order to recreate the remaining GMC encounters in \textsc{nbody6tt}.}
 \label{fig:GMC_removed}
\end{figure} 

This procedure of recreating the orbit is straightforward if the cluster only experiences one strong GMC encounter at a time. If multiple GMC encounters occur in a short period of time, then it is not always possible to recreate the orbit due to the chaotic nature of the many-body problem. We consider that a GMC encounter is successfully recreated when the point of closest approach between the cluster and GMC has an error less than 10 per cent. The impact parameter of the encounter has the highest impact on the energy injected into the cluster (see Eq. \ref{eq:dE}) which is why we base our criterion of a successful orbital recreation on this parameter. The integration time step was initially set to 0.01\,Myr, but we decreased it to $10^{-4}\,$Myr to get acceptable initial conditions for the GMCs during time periods with multiple encounters. If it was not possible to recreate the orbits for all the GMC encounters during a specific period, the GMC which would inject  the least amount of energy into the cluster according to Eq.~\ref{eq:dE} was removed. This process continued in an iterative fashion until all remaining GMC orbits were able to be recreated. 

To test the likely consequences of removing encounters from our analysis, we compare how much energy the removed GMCs would have injected into the cluster compared to the most energetic GMC encounters that occur in the same period of time. We define the energy that would have been injected into the cluster by a GMC that has been removed as $\delta_{E_{\rm{removed}}}$, and compare to the maximum energy injected by a single GMC during the same time period, $\delta_{E_{\rm{max}}}$. The ratio $\delta_f = \delta_{E_{\rm{removed}}}/\delta_{E_{\rm{max}}}$ is plotted in Fig.~\ref{fig:GMC_removed} (top panel) as a function of $\delta_{E_{\rm{removed}}}$ for the C$_{\rm{R}}$ clusters that survive. The majority of clusters have a low $\delta_f$, which means that a much more energetic GMC encounter is dominating the evolution of the cluster during this time. Similarly, most of the removed GMCs have low $\delta_{E_{\rm{removed}}}$ values and should not significantly affect the evolution of the cluster. The only really problematic removals have $\delta_{E_{\rm{removed}}} \geq 0.8$ and $\delta_f \geq 0.6$ which area is represented by the dotted lines. These encounters correspond to 22 GMCs encountering 21 unique clusters; out of the 21 clusters, 6 of them have $M_{\rm c}<50\,$\Mo and are therefore not part of our population analysis. This leaves 15 badly affected clusters, which should not make a significant difference to our total sample of 5748 surviving clusters with masses above $50$ \MO.  The lower panel of Fig.~\ref{fig:GMC_removed} shows the distribution of surviving and destroyed clusters as a function of $\delta_f$. In total, 23\,652 GMC encounters had to be removed from 7534 unique clusters. The vast majority of these GMC encounters are connected to the 6320 out of 7534 clusters that do not survive at the end of their evolution. For the 1214 surviving clusters, a total number of 2341 GMCs had to be removed.

\section{Results}
\label{sec:Results}
We simulate the evolution of 20\,692 individual clusters with masses in the range $[50-24000]$ \MO. The birth times of the clusters are distributed uniformly in the last 1 Gyr. Out of the 20\,692 clusters, 7428 survive in the C$_{\rm{R}}$ simulations and 9903 survive in the  C$_{\rm{N}}$ simulations. We first compare these populations to the observations of \citetalias{Hunt2024}, and then focus on the differences between the C$_{\rm{R}}$ and C$_{\rm{N}}$ clusters to investigate the impact that GMCs have on the clusters' masses, numbers of stars, survivability, and morphology.

\subsection{Mass function}
To compare our cluster population to \citetalias{Hunt2024} we apply their estimated completeness limit (Eq. 6 in \citetalias{Hunt2024}), $R_{100 \%}$, which is given by
\begin{equation}
R_{100 \%} = \left\{ \begin{array}{lr}
  \alpha \, \mathrm{log}_{10}(M /\mathrm{M}_{\odot}) + K  \quad \quad          R_{100 \%} < R_{\rm{break}}\\
  R_{\rm{break}} \quad \quad \quad \quad \quad \quad \quad \quad R_{100 \%} \geq R_{\rm{break}} 
\end{array} \right.,
\label{eq:com100}
\end{equation}
and is valid for cluster masses in the range $40 \le M \le 10^4$ \MO. Here, $\alpha = 633.1$ pc, $K = -1582.6$ pc, and $R_{\rm{break}} = 2792.9$ pc. We only include \citetalias{Hunt2024} clusters which have $R_d \le R_{100 \%}$, where $R_d= \sqrt{X^2 + Y^2}$ is the two dimensional distance from the Sun in the Galactic plane. These clusters are then normalised by the surface area defined by the completeness limit, $\pi R_{100 \%}^2$, in order to compare the mass and age functions of the observations with our simulated cluster populations.

\begin{figure*}
 \includegraphics[width=1.0\textwidth]{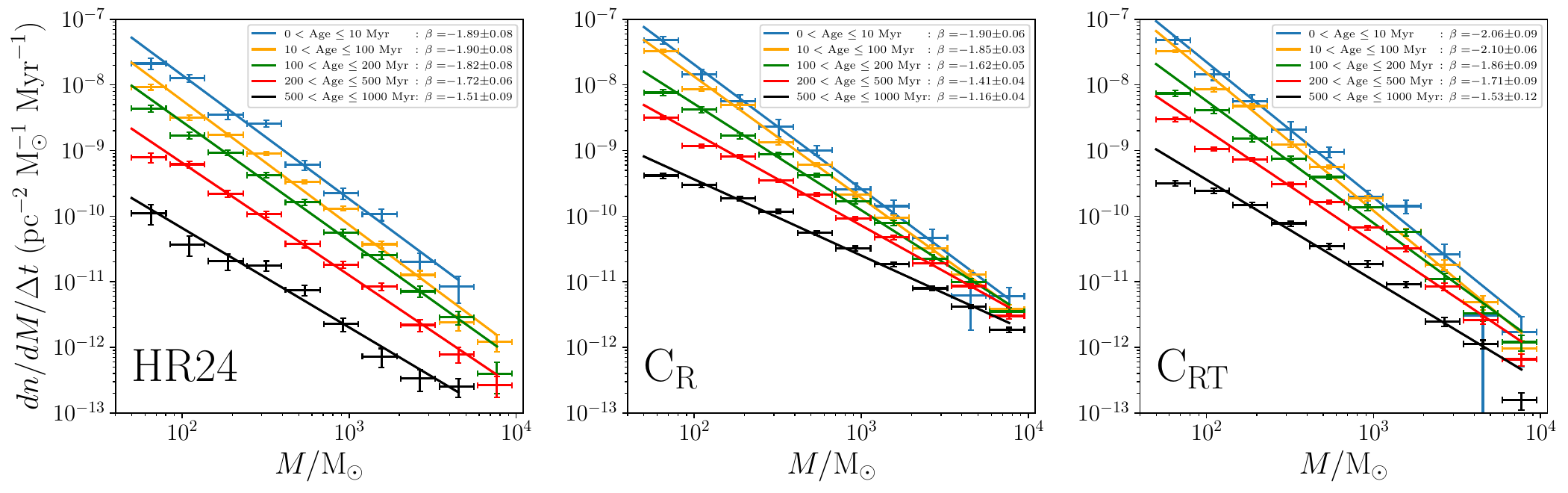}
 \caption{Mass functions of the observed and simulated populations binned over different age ranges. {\it Left:} Observed clusters taken from HR24.  {\it Middle:} Simulated clusters including the effects of GMCs.  {\it Right:} Simulated clusters including the effects of GMCs and a truncated ICMF. Points show binned data in different age bins: top to bottom the bins are $0-10\,$Myr (blue); $10-100\,$Myr (orange); $100-200$\,Myr (green); $200-500$\,Myr (red), and 500\,Myr to 1\,Gyr (black). Lines show best-fit single power-laws.}
\label{fig:MF_dt}
\end{figure*}

We first look at the mass function, which we divide into different age ranges in order to investigate its evolution with time. The mass functions of the \citetalias{Hunt2024} and C$_{\rm{R}}$ clusters are plotted to the left and middle in Fig.~\ref{fig:MF_dt}.  We fit each age bin with a single power-law; the slopes $\beta$ are listed in Table \ref{table:MF_Age}. For clusters younger than 100 Myr we see good agreement between the C$_{\rm{R}}$ clusters and observations. For ages between 100 and 200\,Myr there is less change in the mass function of the observations, with $\beta = -1.82\pm 0.08$, whereas for the C$_{\rm{R}}$ clusters the slope flattens to $\beta = -1.62\pm 0.05$. As the ages of the clusters increase, so does the discrepancy between observations and simulations. For the oldest bin, with cluster ages between 500\,Myr and 1\,Gyr, the observations have $\beta = -1.51\pm 0.09$, whereas the C$_{\rm{R}}$ clusters have $\beta = -1.16\pm 0.04$. This discrepancy cannot be explained by the fact that GMCs have a too destructive impact on the cluster population, since the slope of the mass function for the oldest C$_{\rm{N}}$ clusters is also much flatter than the observations, with a value of $\beta = -1.28\pm 0.05$.

\begin{table*}
 \begin{tabular*}{1.5\columnwidth}{@{\extracolsep{\fill}} cccccc}
 \hline
Age interval                             &  \citetalias{Hunt2024}    & C$_{\rm{R}}$       &  C$_{\rm{N}}$     &  C$_{\rm{RT}}$  & C$_{\rm{NT}}$  \\
                                         &  $\beta$                  & $\beta$            &  $\beta$          &  $\beta$        &  $\beta$ \\ 
\hline
$0 < \rm{Age} \leq 10$ Myr               &   $-1.89\pm 0.08$         &  $-1.90\pm 0.06$   &  $-1.89\pm 0.06$  & $-2.06\pm 0.09$ & $-2.06\pm 0.09$\\
$10 < \rm{Age} \leq 100$   Myr           &   $-1.90\pm 0.08$         &  $-1.85\pm 0.03$   &  $-1.87\pm 0.03$  & $-2.10\pm 0.06$ & $-2.13\pm 0.07$ \\
$100 < \rm{Age} \leq 200$  Myr           &   $-1.82\pm 0.08$         &  $-1.62\pm 0.05$   &  $-1.72\pm 0.05$  & $-1.86\pm 0.09$ & $-1.96\pm 0.08$\\
$200 < \rm{Age} \leq 500$  Myr           &   $-1.72\pm 0.06$         &  $-1.41\pm 0.04$   &  $-1.52\pm 0.04$  & $-1.71\pm 0.09$ & $-1.79\pm 0.08$\\
$500 < \rm{Age} \leq 1000$ Myr           &   $-1.51\pm 0.09$         &  $-1.16\pm 0.04$   &  $-1.28\pm 0.05$  & $-1.53\pm 0.12$ & $-1.62\pm 0.10$\\
\hline
 \end{tabular*}
 \caption{Coefficients of power-law fits to the mass function for different age ranges and observed / modelled populations.}
 \label{table:MF_Age}
\end{table*}

In addition to the discrepancy between the slopes of the mass functions for older clusters, we also see a difference between the relative number of clusters in each age group. The observed number of clusters in each group varies more rapidly with age compared to our simulations. This suggests that the high-mass clusters in our simulations are losing mass at a slower rate compared to what the observations indicate. This cannot be explained by the strength of the gravitational potential being too weak either, since increasing the rate of tidal disruption would accelerate the dissolution of low-mass clusters more rapidly than for high-mass clusters. 

Observations of GMC mass functions, both in our Galaxy \citep{Rice2016,Miville2017} and other galaxies \citep{Rosolowsky2007, Colombo2014}, show that there is a truncation at high GMC masses. A truncation of the GMC mass distribution has been suggested to be linked with a truncation of the mass of stellar clusters \citep{Kruijssen2014}. This is supported by observations where such a truncation is seen for stellar clusters in several galaxies \citep{Jordan2007,Adamo2015,Hollyhead2016,Adamo2017,Johnson2017, Messa2018}. 

We therefore investigated whether the discrepancy between the model and observations could be an effect of a truncated ICMF.  We took our initial population of 20\,692 clusters and resampled them based on the Schechter \citep{Schechter1976} distribution function with slope $\beta_s = -2.0$ and truncation mass $M_t$,
\begin{equation}
\frac{\ud N}{\ud M} \propto M^{\beta_s}\exp\left(-\frac{M}{M_t}\right).
\label{eqn:Schechter}
\end{equation}
 
\begin{figure}
\centering
 \includegraphics[width=\columnwidth]{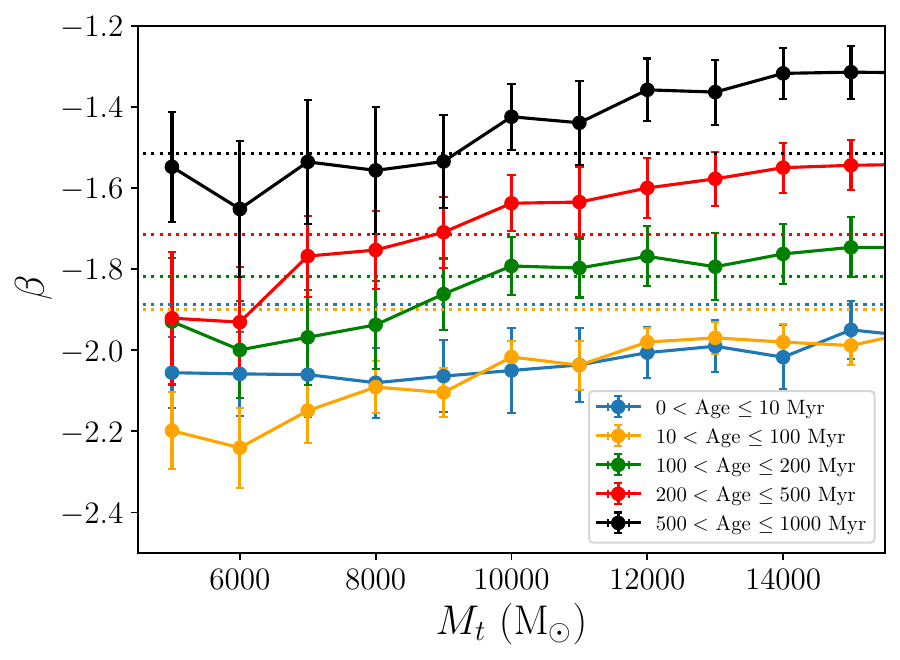}
\caption{Effect of applying truncation to the ICMF on the mass function slopes. The slope, $\beta$, is plotted as a function of the ICMF truncation mass, $M_t$ (see Eq.~\ref{eqn:Schechter}).  The sample of clusters is split by age, into bins (from bottom to top) of $0 - 10$\,Myr (blue); $10-100\,$Myr (orange); $100-200$\,Myr (green); $200-500$\,Myr (red), and 500\,Myr to 1\,Gyr (black).  Dotted lines of corresponding colour show the fits to the observations in the same age bins.}
\label{fig:beta_Mt}
\end{figure}

It is instructive to investigate how the truncation of the ICMF affects the slope of the mass function in different age ranges. This can be seen in Fig.~\ref{fig:beta_Mt}, where the estimated $\beta$ can be seen for each age interval as a function of the truncation mass from which our model clusters are resampled. The horizontal dotted lines show the estimated $\beta$ for the observation for each age interval. For truncation masses above 15\,000 \MO, $\beta$ does not change since most of the cluster distribution is unchanged. However, when the truncation mass goes below 15\,000 \MO, we see a decrease in $\beta$ that is stronger for older clusters. As the truncation mass of the ICMF decreases further, the greater effect it has on the derived value of $\beta$. For truncation masses below $\sim 8\,000$ \MO, the cluster population with ages between 100 and 200 Myr starts to deviate significantly from a slope of $-2$, which is likely a consequence of stochastic effects and the normalisation scheme we use to analyse the whole cluster mass range.

We find that the truncation mass which best replicates the observations is $M_t \sim 9000$\,\Mo which we determine by using a minimum chi-square estimation to each mass and age bin as shown in Figure~\ref{fig:beta_Mt}. We only compare the $\mathrm{C_R}$ clusters to the observations when determining the truncation mass, since the $\mathrm{C_N}$ cluster population does not take into account the tidal effects of the GMCs, and thus is not a realistic representation of the cluster population of the Milky Way. The resulting mass functions which emerge from drawing the cluster population from a truncated ICMF with a truncation mass of $9000$ \Mo are plotted in the right-hand panel of Fig.~\ref{fig:MF_dt}. The mass function slope is slightly too steep for ages below $100$\,Myr, but otherwise there is good agreement between all other age groups and the observations. Henceforth, we adopt an ICMF truncated at $9000$ \Mo for our clusters when comparing to the catalogue of \citetalias{Hunt2024}, and refer to these samples as C$_{\rm{RT}}$ and C$_{\rm{NT}}$ for simulations with and without GMC disruption, respectively. 

\subsubsection{Evolution of the mass function}
\begin{figure}
\centering
 \includegraphics[width=\columnwidth]{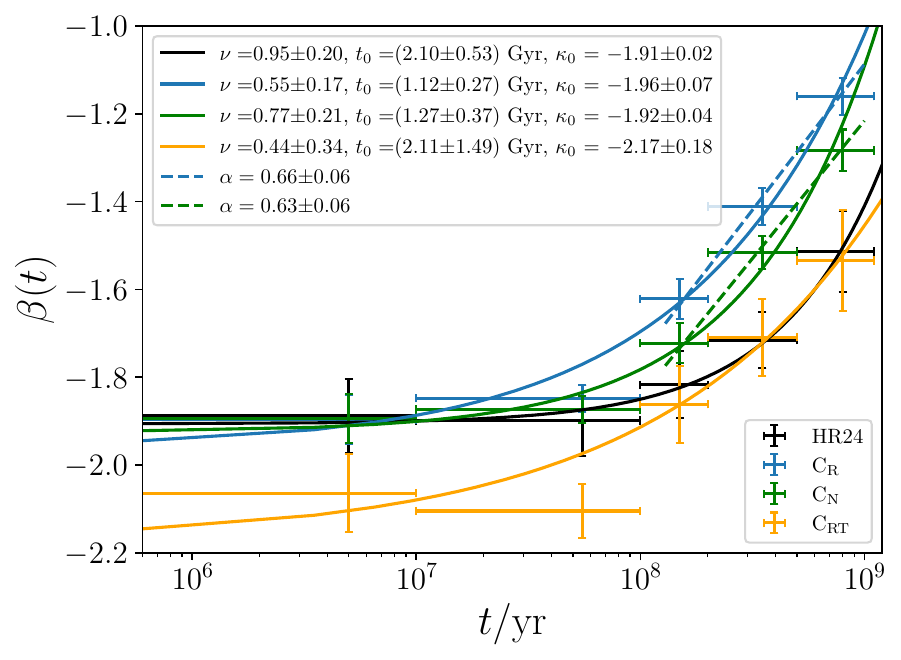}
\caption{The evolution of the mass function slope, $\beta$, as a function of time $t$.  Black points show the observed population from \protect\citetalias{Hunt2024}, blue points our complete model population ($\rm C_R$), green points ($\rm C_N$), and orange points the model population with a truncation mass of 9\,000\,\MO ($\rm C_{RT}$).  The solid and dashed lines show fits of the form given in Eq.~\protect\ref{eq:kappa} and ~\protect\ref{eq:kappa2}, respectively.}
\label{fig:kappa}
\end{figure}

The evolution of the mass function can be investigated by defining the slope of the mass function as a function of time, $\beta(t)$, by using the age-binned mass functions of Fig. \ref{fig:MF_dt}. We adopt a similar functional form to \citetalias{Hunt2024}:
\begin{equation}
\beta(t) = \left ( \frac{t}{t_0} \right )^{\nu} + \kappa_0.
\label{eq:kappa}
\end{equation}
Here $\kappa_0$ is the initial slope of the mass function, $\nu$ is the slope of the power-law, and $t_0$ is the age at which the slope of the initial mass function has increased by $1$. We fit Eq. \ref{eq:kappa} to our distributions and list the parameter estimates in Table \ref{table:kappa}. The evolution of the mass function can be seen in Fig. \ref{fig:kappa} for the C$_{\rm{R}}$, C$_{\rm{N}}$, C$_{\rm{RT}}$, and \citetalias{Hunt2024} clusters. All cluster populations show reasonable agreement with $\kappa_0 \sim -2$ which we expect, since this is the slope of the ICMF. For the observations and the C$_{\rm{RT}}$ clusters, we find $t_0$ values of ($2.10\pm 0.53$) Gyr and ($2.11\pm 1.49$) Gyr, which indicates that the evolution of their mass functions occur over similar time-scales. For C$_{\rm{R}}$ clusters, $t_0 = (1.12\pm 0.27)$ Gyr which means that the mass function starts to evolve significantly earlier.  The rapid evolution of $\beta$ in the C$_{\rm{R}}$ population is driven by the dissolution of low-mass clusters, whereas the high-mass clusters lose mass more slowly and survive much longer. Truncation of the mass function reduces the number of clusters at higher masses and so the mass function flattens more slowly. For the observed population, $\nu = 0.95\pm 0.20$, which suggests nearly linear growth of $\beta$. The values of $\nu$ for the C$_{\rm{RT}}$ and C$_{\rm{R}}$ clusters are lower: $0.44\pm 0.34$ and $0.55\pm 0.17$, respectively. Even so, the late-time evolution of $\beta$ for the observed clusters is very similar to that of the C$_{\rm{RT}}$ population. 

For age intervals older than 100\,Myr, we see a difference between $\beta$ for the C$_{\rm{R}}$ and C$_{\rm{N}}$ clusters with values of $-1.62\pm 0.05$ and $-1.72\pm 0.05$ in the age interval $100-200$ Myr, indicating that GMC encounters start to affect the cluster population in this age range. For clusters older than 100 Myr, the rate of change of the mass function slope is similar for the C$_{\rm{R}}$ and C$_{\rm{N}}$ clusters. We show this in Fig \ref{fig:kappa}, by fitting the three oldest age intervals to the function
\begin{equation}
\beta(t) = \alpha \, \rm{log_{10}}(t/\rm{yr}) + b.
\label{eq:kappa2}
\end{equation}
The slope of $\beta(t)$, i.e. $\alpha = d\beta/d \rm{log_{10}}(t/\rm{yr})$, is $0.66\pm 0.06$ and $0.63\pm 0.06$ for the $\rm{C_R}$ and $\rm{C_N}$ clusters, respectively. This indicates that it is the rest of the Galactic environment that drives the dissolution of clusters at this point, at least for the majority of the cluster population. The same effect is not present between the C$_{\rm{RT}}$ and C$_{\rm{NT}}$ clusters which is a result of the lack of high-mass clusters in the truncated ICMF. The effect is further blurred out because the truncation of the ICMF decreases the number of clusters in our sample, which increases the uncertainties of $\beta$.

\begin{table}
 \begin{tabular*}{\columnwidth}{@{\extracolsep{\fill}} lccc}
 \hline
Name                  &   $\nu$        & $t_0$    &    $\kappa_0$        \\
                      &                & (Gyr)    &                      \\ 
\hline
\citetalias{Hunt2024} & $0.95\pm 0.20$ & $2.10\pm 0.53$ & $-1.91\pm 0.02$ \\
C$_{\rm{R}}$          & $0.55\pm 0.17$ & $1.12\pm 0.27$ & $-1.96\pm 0.07$ \\
C$_{\rm{N}}$          & $0.77\pm 0.21$ & $1.27\pm 0.37$ & $-1.92\pm 0.04$ \\
C$_{\rm{RT}}$         & $0.44\pm 0.34$ & $2.11\pm 1.49$ & $-2.17\pm 0.18$ \\
C$_{\rm{NT}}$         & $0.80\pm 0.51$ & $1.63\pm 1.11$ & $-2.13\pm 0.08$ \\
  \hline
 \end{tabular*}
 \caption{Parameters to fits of the mass function evolution given in Eq. \ref{eq:kappa}.}
 \label{table:kappa}
\end{table}

\subsubsection{Overall mass function}
\begin{figure}
  \includegraphics[width=\columnwidth]{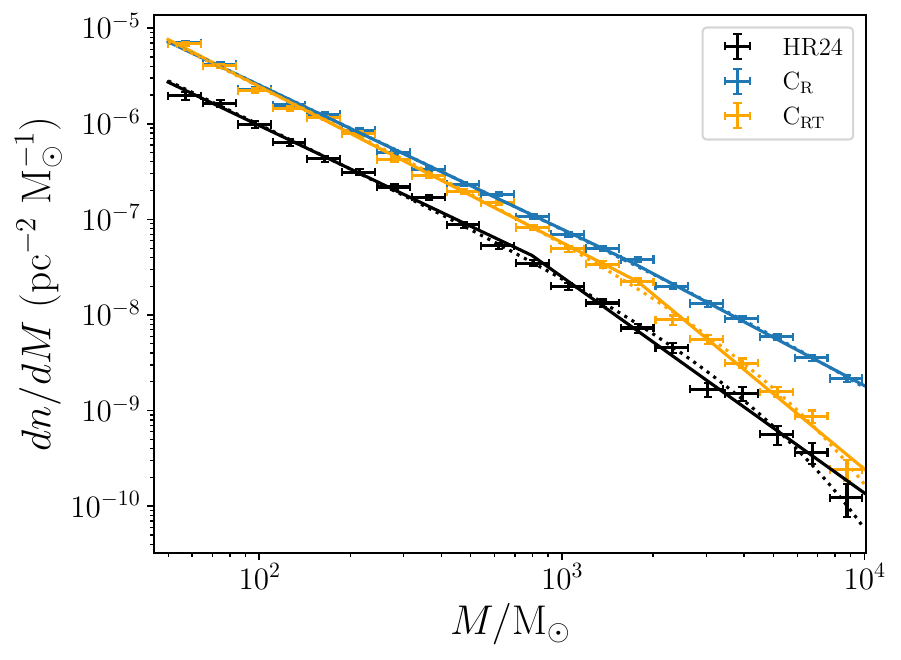}
 \caption{The mass function of the total observed population (HR24, black), the complete model population ($\rm C_R$, blue) and the model population with the ICMF truncated at 9000\,\Mo ($\rm C_{RT}$, orange). Solid lines show broken power-law fits, dotted lines fits with a Schechter function. The vertical difference between the cluster models and observations is a consequence of the normalisation of the modelled population and so the difference is not significant.}
\label{fig:MF_fit}
\end{figure}

Our analysis shows that the evolution of the mass function is strongly affected by whether the ICMF is truncated. This also affects the shape of the mass function of the population of all ages up to 1 Gyr. To analyse the overall mass functions, we fit three different parameterisations: a single power-law with slope $\beta$; a broken power-law with slopes $\beta_1$, $\beta_2$, and break mass $M_{\rm{break}}$; and a Schechter function with slope $\beta_s$ and truncation mass $M_t$. The fits are listed in Table \ref{table:MF_fits}, and plotted in Fig.~\ref{fig:MF_fit}, where the solid lines show the broken power-law fits and the dotted lines the Schechter functions. The C$_{\rm{R}}$ mass function shows minimal evidence for a break and can be well described by a single power-law with $\beta=-1.55\pm 0.02$. The observations and the C$_{\rm{RT}}$ clusters are less well described by a single power-law. Even so, it is worth noting that both mass functions show a similar slope $\beta$ of $\,-1.82\pm 0.05$\,and $\,-1.86\pm 0.05$\,for the observations and C$_{\rm{RT}}$ clusters, respectively.  

Both the \citetalias{Hunt2024} and C$_{\rm{RT}}$ clusters are well described by either a broken power-law or a Schechter function. For the broken power-law fits, the observed break is at a lower mass of $M_{\rm{break}} = (0.79\pm 0.58) \times 10^3$ \MO, compared to the C$_{\rm{RT}}$ clusters at $(1.78\pm 0.58) \times 10^3$ \MO, though these are consistent given the fit uncertainties. The C$_{\rm{RT}}$ mass function has a steeper $\beta_1$ with a value of $-1.63\pm 0.03$ compared to the C$_{\rm{R}}$ clusters which show a $\beta_1 = -1.50\pm 0.02$ which is in better agreement with observations. The C$_{\rm{RT}}$ clusters have a steeper $\beta_2$ of $-2.62\pm 0.09$ compared to the observations with $\beta_2=-2.26\pm 0.06$. 

The Schechter function is a good description for all mass functions which all show similar $\beta_s \sim -1.50$, however, the C$_{\rm{R}}$ and C$_{\rm{N}}$ clusters show no significant truncation, as expected. For the C$_{\rm{RT}}$ clusters, we see a clear truncation with $M_{t} = (4.04\pm 0.49) \times 10^3$ \Mo which is in better agreement with the observations of $(3.52\pm 0.47) \times 10^3$ \MO, compared to the fits of a broken power-law. We therefore find the best agreement between the observations and the C$_{\rm{RT}}$ clusters if their mass functions are described by Schechter functions.

\begin{table*}
 \begin{tabular*}{\textwidth}{@{\extracolsep{\fill}} lccccccccc}
  Name  & Power-law & & Broken Power-law & & & &Schechter    \\
  \hline
  & $\beta$ & $\chi^2_{\nu}$ & $\beta_1$&   $M_{\rm{break}}$ &   $\beta_2$ & $\chi^2_{\nu}$ & $\beta_s$ & $M_{t}$ & $\chi^2_{\nu}$ \\
  &  &   &     & $(10^3\,\rm{M}_{\odot})$ & & & &  $(10^3\,\rm{M}_{\odot})$ & \\
\citetalias{Hunt2024} &  $-1.82\pm 0.05$ & $1.53$ & $-1.51\pm 0.04$   & $0.79\pm 0.58$  &  $-2.26\pm 0.06$ & $0.34$ & $-1.50\pm 0.04$ & $3.52\pm 0.47$ & $0.31$ \\

C$_{\rm{R}}$          &  $-1.55\pm 0.02$  & $0.29$ &  $-1.50\pm 0.02$ &  $1.65\pm 1.77$ &  $-1.68\pm 0.05$ & $0.21$ & $-1.50\pm 0.03$ & $24.7\pm 10.9$ & $0.23$ \\
C$_{\rm{N}}$          &  $-1.60\pm 0.02$  & $0.37$ & $-1.52\pm 0.02$  &  $1.68\pm 1.56$ &  $-1.80\pm 0.05$ & $0.17$ & $-1.50\pm 0.02$ & $14.5\pm 3.7$ & $0.18$ \\
C$_{\rm{RT}}$         &  $-1.86\pm 0.05$  & $1.70$ &  $-1.63\pm 0.03$ &  $1.78\pm 0.58$ &  $-2.62\pm 0.09$ & $0.26$ & $-1.54\pm 0.03$ & $4.04\pm 0.49$ & $0.25$ \\
C$_{\rm{NT}}$         &  $-1.89\pm 0.04$  & $2.35$ & $-1.64\pm 0.03$  &  $1.96\pm 0.74$ &  $-2.88\pm 0.14$ & $0.44$ & $-1.52\pm 0.04$ & $3.37\pm 0.41$ & $0.34$ \\


  \hline
 \end{tabular*}
 \caption{Parameters of fits to the mass function where all cluster ages are included. The fits include a power-law, a broken power-law, and a Schechter function. The reduced chi-squared, $\chi^2_{\nu}$, is listed for each fit.}
 \label{table:MF_fits}
\end{table*}

The mass functions of stellar cluster populations in other spiral galaxies also show signs of truncation. Observations show that in M83 with $M_t = (1.60\pm 0.30) \times 10^5$ \Mo \citep{Adamo2015}; NGC 1566 $M_t = 2.5 \times 10^5$ \Mo \citep{Hollyhead2016}; NGC 628 $M_t = (2.03\pm 0.81) \times 10^5$ \Mo \,\citep{Adamo2017}; and M51 $M_t = (1.00\pm 0.12) \times 10^5$ \Mo \citep{Messa2018}. These truncation masses are higher than for the observed \citetalias{Hunt2024} cluster population. A smaller truncation mass of $\sim 8.5 \times 10^3$ \Mo has been found in M31\citep{Johnson2017}, however, this is still substantially greater than the {\it present day} truncation mass observed in the Milky Way.\footnote{Note that the {\it observed} truncation mass depends not only on the {\it initial} truncation mass but also the star-formation history, since clusters lose mass with time.}

In M83, \citet{Adamo2015} found that $M_t$ decreases significantly from the inner to the outer regions of the galaxy, with values from $4.0 \times 10^5$ to $2.5\times10^4$ \MO. This is also the case in NGC 628 and M51, where $M_t$ decreases from $4.85 \times 10^5$ to $1.04 \times 10^5$ \Mo \citep{Adamo2017} and from $2.51\times 10^5$ to $3.6\times 10^4$ \MO \citep{Messa2018b}, respectively. Furthermore, \citet{Johnson2017} have suggested that $M_t$ could scale with the star formation rate surface density. There is therefore probably an environmental dependence on the cluster formation, which might explain why the observed $M_t$ of $(3.52\pm 0.47) \times 10^3$ \Mo in the Milky Way is so much lower compared to what is observed for other spiral galaxies. The observed cluster population in the Milky Way is far from complete. \citetalias{Hunt2024} estimate that their catalogue only contains $\sim 4$ per cent of the total cluster population of the Milky Way and could further help explain the low value of $M_t$ that is observed. 

A comparison between the C$_{\rm{RT}}$ and C$_{\rm{NT}}$ mass functions suggests that the mass function is ineffectual as a tool to investigate the effects of GMCs: our measured power-law mass function slopes $\beta$ are the same within the uncertainties.  A marginally significant difference is visible between the C$_{\rm{R}}$ and C$_{\rm{N}}$ mass functions for older clusters.

\subsection{Age functions}
\begin{figure}
 \includegraphics[width=1.0\columnwidth]{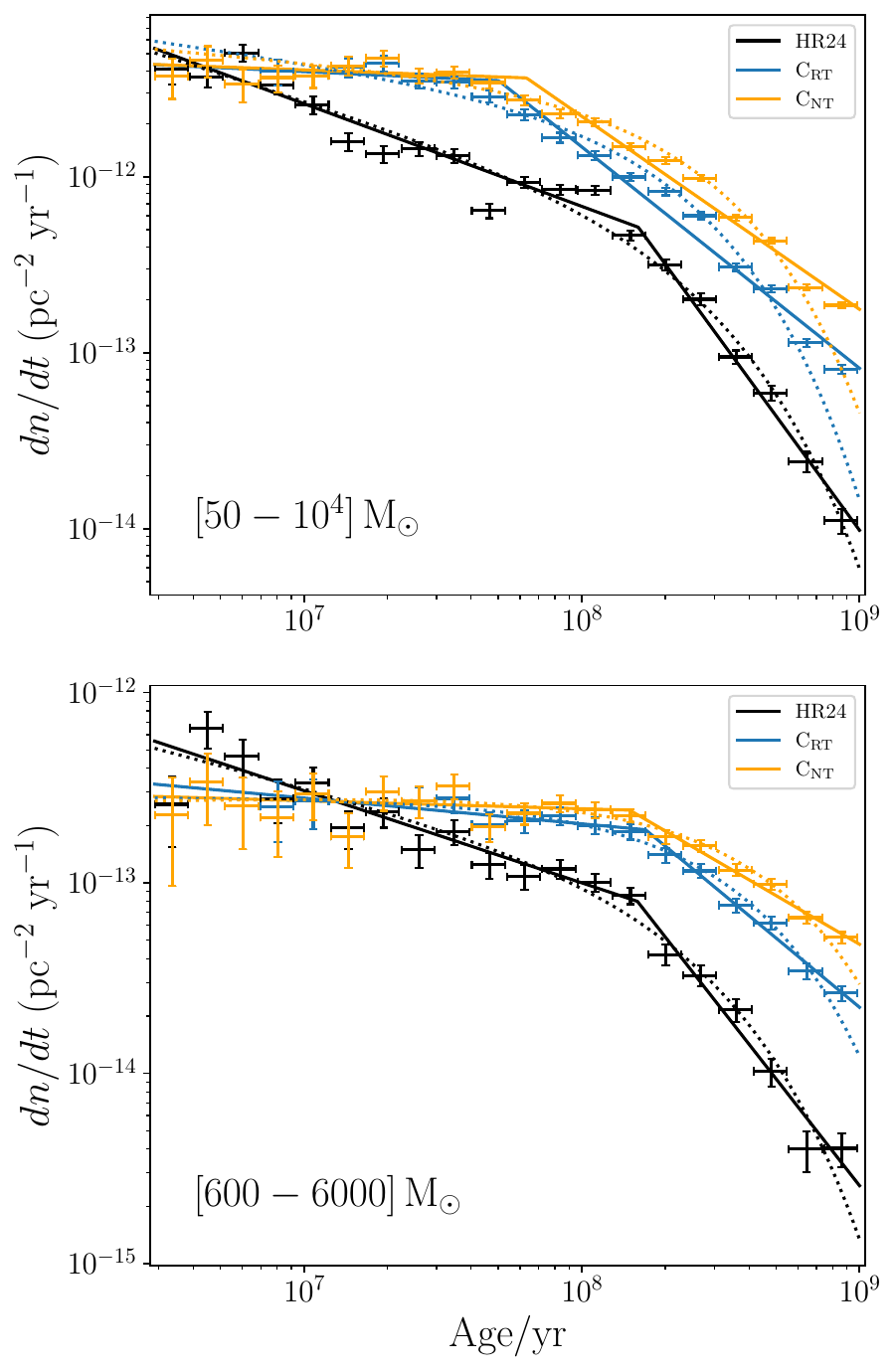}
 \caption{The age distribution of the observed population (black), the model population including the effects of GMC disruption with the ICMF truncated at 9000\,\Mo (blue), and the model population excluding the effects GMCs, also with the ICMF truncated at 9000\,\Mo (orange).  Solid lines show fits with broken power-laws, dotted lines with Schechter functions. {\it Top:} Complete populations. {\it Bottom:} subset of clusters between 600 and 6000\,\MO.
 }
\label{fig:AF_dM_fit}
\end{figure}

The age distributions of the total cluster populations of the \citetalias{Hunt2024}, C$_{\rm{RT}}$, and C$_{\rm{NT}}$ clusters are shown in the top panel of Fig.~\ref{fig:AF_dM_fit}. We fit each age distribution to two different functional forms: a broken power-law with slopes $\gamma_1$ and $\gamma_2$, and break age $t_{\rm{break}}$; and a Schechter function with slope $\gamma_s$ and truncation age $t_s$. The broken power-law and Schechter fits are represented by solid and dotted lines, and are listed in Table \ref{table:AF}. \citetalias{Hunt2024} find that a broken power-law is the best fit to the observed age function; we concur and find the same result for our model C$_{\rm{RT}}$ and C$_{\rm{NT}}$ clusters. 

\begin{table*}
 \begin{tabular*}{\textwidth}{@{\extracolsep{\fill}} lcccccccc}
  Name & Broken power-law &  & & &Schechter    \\
  \hline
 & $\gamma_1$&   $\rm{log}_{10}(t_{\rm{break}}/\rm{yr})$ &   $\gamma_2$  & $\chi^2_{\nu}$ & $\gamma_s$ & $\rm{log}_{10}(t_{s}/\rm{yr})$ & $\chi^2_{\nu}$ \\
 \hline
 Mass range:$[50 - 10^4]$ \Mo \\

\citetalias{Hunt2024}  &  $-0.58\pm 0.05$ & $8.20\pm 0.12$ & $-2.16\pm 0.22$ & $2.69$ & $-0.49\pm 0.07$ & $8.41\pm 0.09$ & $3.02$ \\

C$_{\rm{R}}$ &  $-0.07\pm 0.05$ & $7.67\pm 0.05$ & $-1.07\pm 0.03$ & $0.87$ & $-0.24\pm 0.06$ & $8.41\pm 0.08$ & $1.95$ \\
C$_{\rm{N}}$ &  $-0.04\pm 0.05$ & $7.77\pm 0.09$ & $-0.95\pm 0.04$ & $1.04$ & $-0.14\pm 0.05$ & $8.46\pm 0.07$ & $1.68$ \\

C$_{\rm{RT}}$ &  $-0.07\pm 0.06$ & $7.70\pm 0.12$ & $-1.25\pm 0.05$ & $1.10$ & $-0.22\pm 0.06$ & $8.33\pm 0.06$ & $1.60$ \\
C$_{\rm{NT}}$ &  $-0.06\pm 0.05$ & $7.80\pm 0.05$ & $-1.10\pm 0.05$ & $1.16$ & $-0.13\pm 0.05$ & $8.40\pm 0.06$ & $1.39$ \\

  \hline
Mass range:$[600 - 6000]$ \Mo \\
\citetalias{Hunt2024}     &  $-0.48\pm 0.04$ & $8.20\pm 0.08$ & $-1.87\pm 0.13$ & $0.91$ & $-0.38\pm 0.06$ & $8.42\pm 0.07$ & $1.24$ \\
C$_{\rm{R}}$     &  $-0.14\pm 0.03$ & $8.24\pm 0.12$ & $-0.91\pm 0.06$ & $0.68$ & $-0.04\pm 0.04$ & $8.66\pm 0.06$ & $0.63$ \\
C$_{\rm{N}}$     &  $-0.06\pm 0.04$ & $8.16\pm 0.13$ & $-0.62\pm 0.05$ & $0.83$ & $-0.02\pm 0.04$ & $8.79\pm 0.07$ & $0.83$ \\

C$_{\rm{RT}}$    &  $-0.14\pm 0.03$ & $8.23\pm 0.10$ & $-1.21\pm 0.06$ & $0.40$ & $-0.01\pm 0.04$ & $8.51\pm 0.04$ & $0.43$ \\
C$_{\rm{NT}}$    &  $-0.04\pm 0.04$ & $8.17\pm 0.10$ & $-0.85\pm 0.05$ & $0.59$ & $-0.03\pm 0.05$ & $8.62\pm 0.06$ & $0.68$ \\

 \hline
  
 \end{tabular*}
 \caption{Age function fits to a broken power-law and a Schechter function. {\it Top:} Complete populations.  {\it Bottom:} subset of clusters between 600 and 6000\,\MO.}
 \label{table:AF}
\end{table*}

For a population with no cluster dissolution we would expect to see a flat age function, and indeed the age functions of both the C$_{\rm{RT}}$ and C$_{\rm{NT}}$ models are flat up until the break at $\sim 50$ Myr. Thereafter, the C$_{\rm{RT}}$ and C$_{\rm{NT}}$ populations have $\gamma_2=-1.25\pm 0.05$ and $\gamma_2=-1.10\pm 0.05$, respectively. The presence of GMCs decreases the number of older clusters which causes a shift between the ages functions of the C$_{\rm{RT}}$ and C$_{\rm{NT}}$ populations which occur at ages older than $\sim 70$ Myr. Even though there are fewer clusters in the C$_{\rm{RT}}$ population, the slope of $\gamma_2$ remains relatively unchanged. The non-truncated population has a very similar age function but with a slightly less steep power-law for older clusters, $\gamma_2=-1.07\pm 0.03$, since the total population is dominated by lower-mass clusters, except for the oldest cluster ages.

Similarly to the modelled clusters, the age function for the observations is also best described by a broken power-law. However, unlike our simulations, the observations show a decline in the number of clusters which starts at young ages after $\sim 1$ Myr and have $\gamma_1=-0.58\pm 0.05$. The observations therefore suggest that the dissolution of clusters is already ongoing from the earliest of times. The observed break in the age function is located at a later age compared to our models, at $\sim 160$ Myr, followed by a decline with  $\gamma_2=-2.16\pm 0.22$, which is significantly steeper than in our simulations.

In order to investigate the discrepancy between the observed and modelled age functions, we also computed the age function over a subset of cluster masses. Since the overall age function is dominated by low-mass clusters for which the observations are more likely to be incomplete, it is useful to consider a higher mass interval. The age functions for clusters between 600 and 6000\,\Mo can be seen in the bottom of Fig.~\ref{fig:AF_dM_fit}. For this mass range, the break ages in the models and observations are both $\sim 160$ Myr. For the observations, the slope for young clusters $\gamma_1=-0.48\pm 0.04$ which is slightly less steep than for the full population, whereas the slope for the C$_{\rm{RT}}$ clusters becomes slightly steeper with $\gamma_1=-0.14\pm 0.03$. This is not a physical effect, but rather a stochastic one, since we are using a subsample of our model clusters. Nevertheless, the age functions of our cluster models and observed populations are still very different.

For the higher-mass subset, the observations are quite stochastic, which when combined with the age uncertainty of the \citetalias{Hunt2024} population could contribute to the discrepancies we see. We investigated this by perturbing the ages of our simulated clusters using uncertainties derived from the \citetalias{Hunt2024} population. Each simulated cluster was assigned an age uncertainty corresponding to a \citetalias{Hunt2024} cluster of similar age. We then calculated 100 realisations of the age function in which the new ages of the C$_{\rm{R}}$ clusters were sampled from Gaussian distributions using the adopted uncertainties for each cluster.  To avoid a truncation at older ages, we added a population of synthetic clusters that were older than 1 Gyr which was based on the original age function. We found that the age uncertainties do not explain the discrepancies between models and observations. For the $600-6000\,$\Mo mass interval we found no change in $\gamma_1$. The biggest difference was seen in the total population where the population-sampled $\gamma_1=-0.21\pm 0.01$, still much flatter than the observed value of $-0.58\pm 0.05$.     

The early disruption of clusters seen in the observations indicate that some of them might be born supervirial, while our cluster models are all born in virial equilibrium. If there is significant gas expulsion during the formation of the cluster, the stars in the cluster can become unbound and will lead to the cluster dissolving on a dynamical time-scale corresponding to a few Myr. Several young clusters in the Milky Way have been observed to be supervirial \citep{Bravi2018, Kuhn2019, Pang2021} and this might explain why we see a discrepancy in the age function between the observations and our model clusters.

\begin{figure}
    \centering
    \includegraphics[width=1.0\linewidth]{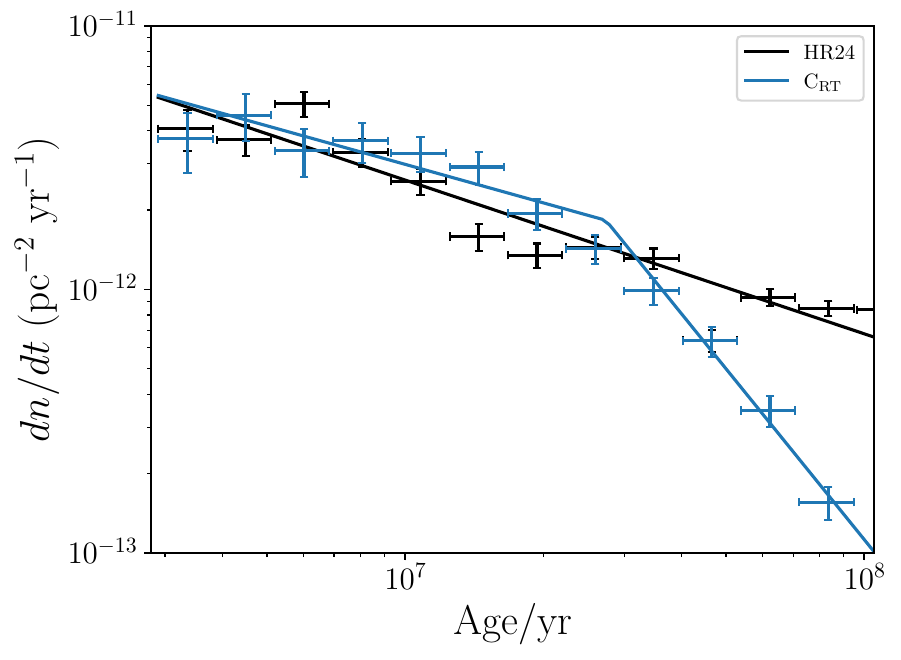}
    \caption{The age function for the C$_{\rm{RT}}$ clusters younger than $100$ Myr which have been re-simulated in \textsc{nbody6tt}. Instead of being born in virial equilibrium, the clusters are born with a viral ratio of $Q = 1.25$.}
    \label{fig:Age_rerun}
\end{figure}

To further investigate whether the discrepancy in the age function during the first 100 Myr could be explained by clusters being born in a higher virial state, we reran $N$-body simulations of the surviving C$_{\rm{RT}}$ clusters with final masses above 50 \Mo and ages younger than 100 Myr. We initially tested $Q=0.75$ and re-fit a broken power-law age function for clusters younger than 100\,Myr. The resulting age function had a slope of $\gamma_1 = 0.06\pm 0.06$, similar to the clusters in virial equilibrium. 
We therefore made another rerun with $Q=1.25$ which can be seen in Fig. \ref{fig:Age_rerun}. This resulted in an age function slope of $\gamma_1 = -0.49\pm 0.09$ which agrees much better with the observations of $\gamma_1 = -0.58\pm 0.05$. The breakpoint occurs at $\rm{log}_{10}(t_{\rm{break}}/\rm{yr}) = 7.44\pm 0.02$, significantly earlier than the observations. However, the post-break slope of $\gamma_2 = -2.15\pm 0.22$ corresponds well with the observed $\gamma_2=-2.16\pm 0.22$. Changing the cluster mass interval for the supervirial born clusters to $600-6000$ \Mo no longer aligns the break point with the observations, as it does for the C$_{\rm{RT}}$ population born in virial equilibrium. This could suggest that there is a correlation between the initial virial ratio and the initial cluster mass.

Clusters that are born with a higher initial mass might be better at converting their primordial gas into stars which reduces the amount of gas expulsion driven by feedback and thereby causing the higher-mass clusters to be born in virial states that are closer to virial equilibrium, i.e. they are more gravitationally bound \citep{Kruijssen2026}. Our investigation of the youngest clusters supports the idea that lower-mass clusters are born more supervirial, which will increase the fraction that are disrupted at early times. However, it is not clear from this experiment what the distribution of the initial virial states should be, and further investigation is needed.


\subsection{Cluster radii}
We now compare our simulations to the observed distributions of the cluster half-number radii, $r_{50}$, the tidal radii $r_t$, and the ratio $r_{50}/r_t$ as functions of age and mass. The ratio $r_{50}/r_t$ gives an estimate of how much the cluster is filling its tidal radius, analogous to the concentration of the cluster. We restrict our comparison to the observed clusters which \citetalias{Hunt2024} classify as high-quality in order to ensure a reliable comparison between the parameters, and, as previously, only consider clusters within $R_{100 \%}$.

\begin{figure*}
    \centering
    \includegraphics[width=1.0\linewidth]{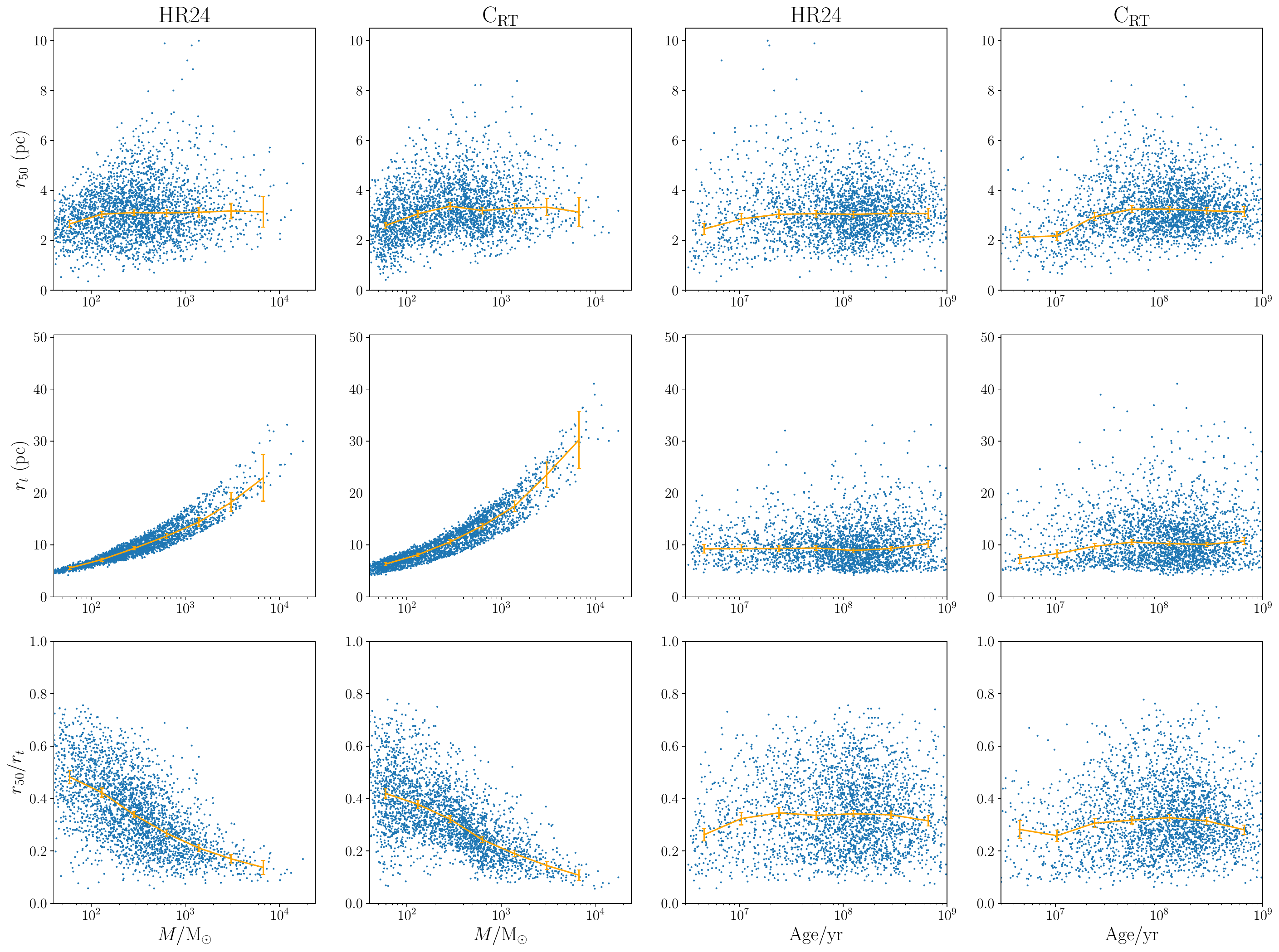}
    \caption{The distribution of $r_{50}$, $r_t$, and $r_{50}/r_t$ as a function of cluster masses and ages for the high-quality clusters of \citetalias{Hunt2024} and the corresponding subsample of C$_{\rm{RT}}$ clusters. A median trend line is shown in orange for each cluster population.}
    \label{fig:C_correlation}
\end{figure*}

In \citetalias{Hunt2024}'s catalogue the lowest stellar mass that can be detected increases with the clusters' two dimensional distance from the Sun in the Galactic plane, $R_d$. We therefore apply a minimum stellar mass cut, $m_{\rm{min}}(R_d)$, when calculating $r_{50}$ for our simulated clusters. We estimated $m_{\rm{min}}(R_d)$ based on the high-quality clusters of \citetalias{Hunt2024}, where we binned the clusters as a function of $R_d$ with a bin size of 100 pc. For each bin, the median minimum stellar mass was calculated to obtain $m_{\rm{min}}(R_d)$ and we used a linear interpolation to estimate the minimum mass cut for each cluster. \citetalias{Hunt2024} clusters with $R_d \lesssim 450$ pc all have $m_{\rm{min}} = 0.09$ \Mo which then increases approximately linearly to a maximum of $m_{\rm{min}} = 0.7$ \MO. We find that if we do not correct for this selection effect, the estimated $r_{50}$ for our clusters is on average approximately 20 per cent larger, since in the models lower-mass stars are, on average, at larger distances from the cluster centres.

The high-quality clusters of \citetalias{Hunt2024} represent a subset of the total cluster population and it is therefore more reasonable to compare them to a similar subset of our clusters. We therefore selected C$_{\rm{RT}}$ clusters with masses and ages which corresponded best to the observed age, $A_{\rm{obs}}$, and mass, $M_{\rm{obs}}$, of each high-quality cluster of \citetalias{Hunt2024}. This was done by selecting model clusters with minimum values of $\Delta p$ which we define as 
\begin{equation}
\Delta p = \left ( \frac{\Delta A}{A_{\rm{obs}}} \right )^2 + \left ( \frac{\Delta M}{M_{\rm{obs}}} \right )^2, 
\end{equation}
where $\Delta A$ and $\Delta M$ are the difference in age and mass between model clusters and each observed cluster.

The first two columns in Fig. \ref{fig:C_correlation} show $r_{50}$¸ $r_t$, and $r_{50}/r_t$ as functions of the cluster mass for the high-quality \citetalias{Hunt2024} clusters and the corresponding subsample of C$_{\rm{RT}}$ clusters. A trend line of binned medians is shown in orange for each population. The radii $r_{50}$ show no significant trend with mass for either observed or modelled clusters. The tidal radii $r_t$ show a strong correlation with mass for both populations, since $r_t$ is calculated directly from the cluster mass according to Eq.~\ref{eq:rt}. The C$_{\rm{RT}}$ clusters show a larger spread in $r_t$ for clusters with masses lower than a few $100$ \MO. This is because the observed clusters are affected by their completeness limit and are therefore generally located at closer distances, whereas our model clusters are not affected by distance. The model clusters therefore experience a larger variation in tidal environments, which is further enlarged by our Milky Way potential not being axisymmetric and causes a larger spread in $r_t$. Low-mass clusters are more sensitive to the galactic tidal field and are therefore also the most tidally filled, both in observations and models.

\begin{figure*}
 \includegraphics[width=0.9\textwidth]{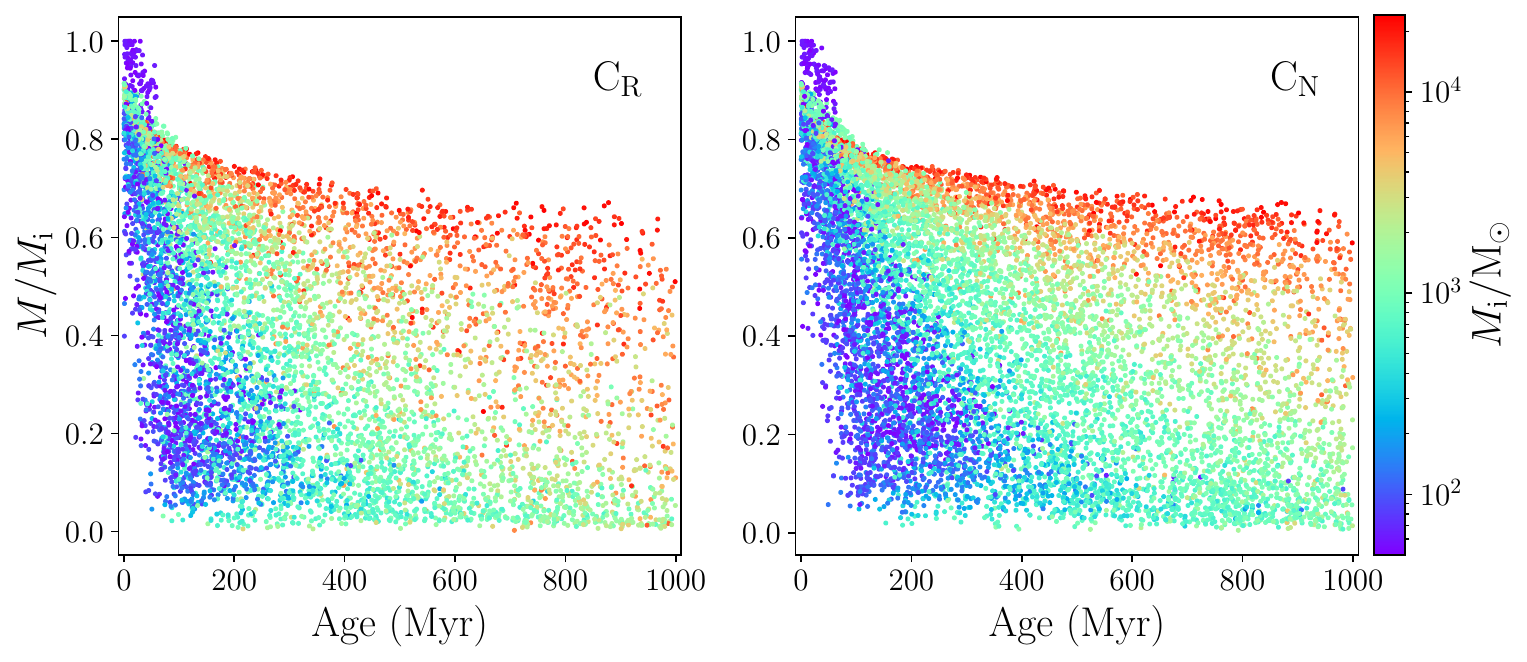}
  \caption{ The fraction of remaining mass, $M/M_{\rm i}$, is plotted as a function of age for model C$_{\rm{R}}$ and C$_{\rm{N}}$ clusters shown to the {\it left} and {\it right}, respectively. Colours show the initial cluster masses.}
 \label{fig:M_age}
\end{figure*} 

The last two columns in Fig. \ref{fig:C_correlation} show $r_{50}$¸ $r_t$, and $r_{50}/r_t$ as a function of the cluster age for observations and models, respectively. The observations show little evolution of $r_{50}$, whereas there is a slight increase at young ages for the C$_{\rm{RT}}$ clusters. This is likely caused by the initial conditions leading to an expansion on a dynamical time-scale, which corresponds to a few Myr. Both observations and models show a flat trend in $r_t$ with a constant median of approximately 10\,pc with similar trends in the degree of tidal filling. For the distribution of radii of the clusters, we generally see good agreement between our models and observations, both in terms of cluster masses and ages.

An interesting extension to the population modelling that we have carried out here would be a ``rewind'' solution, where the initial parameters of the observed clusters are obtained by finding model clusters with similar parameters to the observed clusters and adopting the initial parameters of the modelled clusters. A similar approach was adopted for M83 by \citet{Webb2021}. From this, it could then be possible to estimate an initial mass and size function for the observed cluster population. The initial cluster mass range investigated by \citet{Webb2021} spans the range $15\,000 - 45\,000$ \MO, which means that their clusters are less affected by the galactic tides of M83 compared to the clusters we investigate in the Milky Way. The model clusters in \citet{Webb2021} all follow circular orbits, whereas our clusters can radially migrate. Furthermore, our Galactic model includes structures such as a bar, spiral arms and GMCs which all will affect the tidal histories of each unique cluster. The tidal history is very important when it comes to the shaping the final cluster parameters, especially of low-mass clusters, and it is therefore much more complicated to create a ``rewind'' solution for the Milky Way. It might be possible to do it for the high-mass model clusters, since these will be less affected by the Galactic tides, and is something which could be explored in a future study.

\section{Effects of giant molecular clouds on cluster disruption}
\label{sec:GMC_TT}
When comparing the mass and age functions of the C$_{\rm{R}}$ (realistic, including disruption by GMCs) and C$_{\rm{N}}$ (without effects of GMCs) populations, there is not a great difference. This is largely because not all clusters experience strong encounters with GMCs. 

By combining analytical expressions for cluster mass loss caused by the effects of stellar evolution, tidal stripping, shocking by spiral arms, and encounters with GMCs, \citet{Lamers2006} have predicted the survival time of of initially bound star clusters in the Solar Neighbourhood. One of their conclusions was that GMCs have a large impact on the survival time of lower-mass clusters compared to the mass loss caused by other effects, however, they do not do a direct comparison of the cluster age function with and without the effect of GMCs. The dissolution rate of open clusters in the Milky Way has further been investigated by \citet{Almeida2025} using a similar approach and they conclude that the disruption time is two times longer than previous estimates. Using $N$-body simulations, \citet{Webb2019,Webb2024} constructed a theoretical model for shock-driven cluster mass loss. By comparing statistically identical tidal histories of clusters, they conclude that clusters which experience high-frequency shocks have more similar mass loss histories than clusters experiencing low-frequency shocks. All of the mentioned studies try to account for the global tidal field which affects and dictates the dissolution time of stellar clusters. However, they do not investigate the specific impact that the GMCs have on the tidal field which affect the clusters. In this section, we investigate exactly how the GMCs affect the clusters by comparing individual C$_{\rm{R}}$ and C$_{\rm{N}}$ clusters.

Fig. \ref{fig:M_age} shows the final masses of the clusters as a fraction of their initial birth masses, $M_{\rm{i}}$, as a function of their ages. The clusters are colour-coded according to their initial birth masses. As expected, high-mass clusters retain their mass for longer because their mass loss is less affected by the Galactic tidal field. Most of the clusters with $M_{\rm i} > 10^4$\,\Mo still retain over half of their initial mass after 1\,Gyr. Clusters that are born with masses of $\sim 10^3$\,\Mo barely survive to 1\,Gyr and dominate the lower part of the figure, i.e. $M/M_{\rm{i}} < 0.1$ for ages over 300\,Myr. The C$_{\rm{N}}$ clusters show a similar trend, but with less mass loss per cluster on average and a lower spread in mass loss as a function of initial cluster mass. The reduced mass loss is especially pronounced for clusters with $M_{\rm i}\lesssim300$\,\MO, which reach higher ages of a few hundred Myr in the absence of GMCs.

\begin{figure}
 \includegraphics[width=\columnwidth]{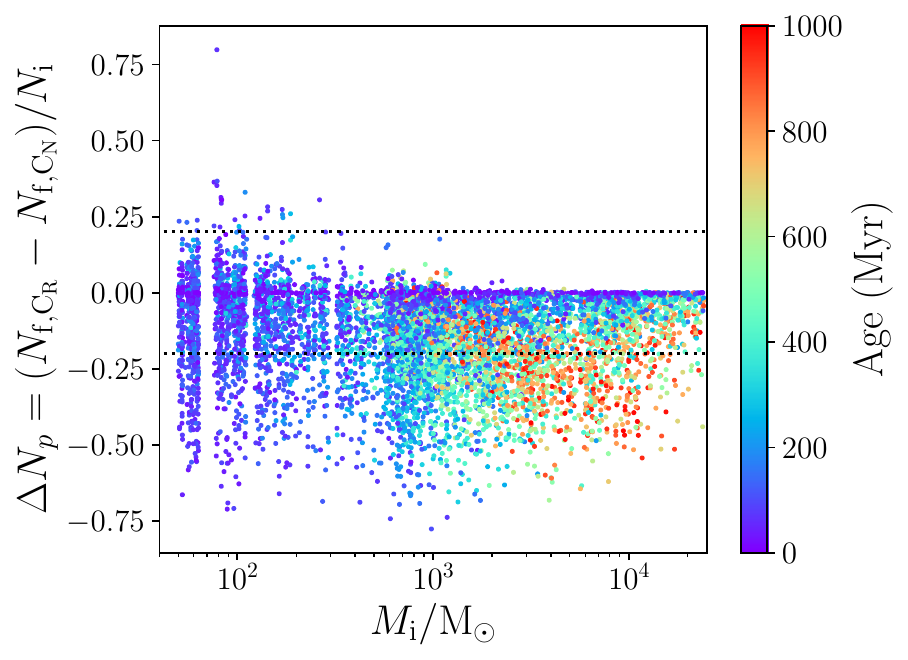}
\caption{The difference between the final number of stars for individual model clusters with and without GMC encounters as a fraction of the initial number of stars, $\Delta N_p$, is plotted as a function of initial cluster mass $M_{\rm i}$.  Points are coloured according to cluster age.  
}
\label{fig:dNp}
\end{figure}

Since a cluster follows the same trajectory through our simulated Milky Way regardless of the presence or absence of GMCs, we can compare individual pairs of clusters between the $\rm C_R$ and $\rm C_N$ simulations. 
We define 
\begin{equation}
\Delta N_p = (N_{\rm{f},\rm{C}_{\rm{R}}} - N_{\rm{f},\rm{C}_{\rm{N}}})/N_{\rm{i}}
\label{eqn:dNp}
\end{equation}
as the difference between the final number of stars $N_{\rm{f}}$ for the C$_{\rm{R}}$ and C$_{\rm{N}}$ clusters divided by the initial number of stars $N_{\rm{i}}$.  We chose to look at the number of stars instead of the cluster mass in order to reduce the effects of stellar evolution.  Fig.~\ref{fig:dNp} shows $\Delta N_p$ as a function of initial cluster mass for each cluster that survives in the C$_{\rm{R}}$ population. The majority of surviving clusters show a small difference in the number of stars, with $|\Delta N_p| < 0.2$. Most of the very young clusters have $\Delta N_p \sim 0$ since they have not had time to evolve in the Galactic tidal field. However, for young clusters there is a large spread in $\Delta N_p$ at lower masses, which indicates that they can quickly lose a large fraction of their stars if they have GMC encounters early in their evolution. In general $\Delta N_p<0$; i.e.~the presence of GMCs increases cluster mass loss, as expected.

There is a clear dependence of $\Delta N_p$ on the initial cluster mass and age. Low-mass clusters are more sensitive to GMC encounters, whereas the higher-mass clusters only show a strong difference in $\Delta N_p$ when old. A few clusters have $\Delta N_p > 0.2$; that is, they lose more stars in the {\it absence} of GMCs. When investigating the cluster population in M51, we found a similar result, and concluded that these clusters are probably ``protected'' by GMCs, in the sense that the tidal field produced by a GMC encounter counteracted the larger-scale galactic tidal field in M51 \citep{Jorgensen2025}. As a consequence, this caused a temporary reduction in cluster mass loss and it is likely the same effect we see here. All clusters with $\Delta N_p > 0.2$ are younger than 300\,Myr. This makes sense since the tidal effects of a GMC can only protect a cluster against the rest of the Galactic tidal field temporarily and in very rare circumstances.

\begin{figure}
 \includegraphics[width=\columnwidth]{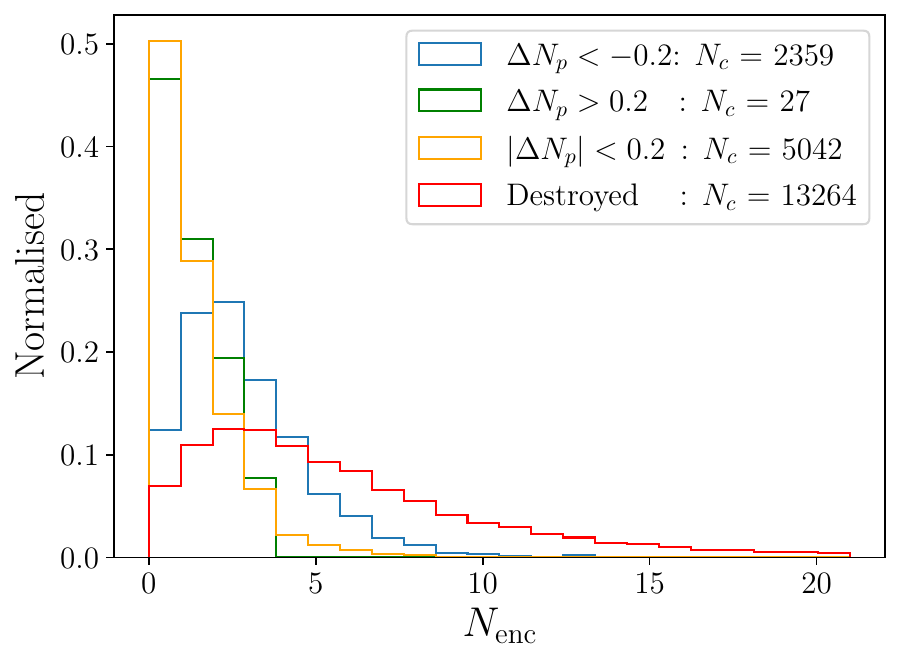}
 \caption{ The distribution of the number of strong encounters with GMCs per cluster, $N_{\rm enc}$. The population is split into four groups according to their values of $\Delta N_p$ (see Eq.~\ref{eqn:dNp}).  Blue bars show clusters with $\Delta N_p<-0.2$ (significant destruction by GMCs); green bars clusters with $\Delta N_p>0.2$ (protection by GMCs); orange bars clusters with $\left|\Delta N_p\right|<0.2$ (limited effect of GMCs) and red bars (destroyed clusters).  The histograms are normalised for each group and the corresponding total number of clusters in the group, $N_c$, is listed in the legend.
 }
 \label{fig:Nenc}
\end{figure}

Fig. \ref{fig:Nenc} shows the distribution of the number of strong GMC encounters that a cluster has encountered, $N_{\rm{enc}}$. Clusters are grouped according to $\Delta N_p$ and strong encounters are defined in the same way as in Section~\ref{sec:GMC_sim}; i.e. only encounters with pericentre separation less than 60\,pc and $\delta_E \ge 0.01$ are included. The median of clusters with $\Delta N_p > -0.2$ (green and orange) have had no strong GMC encounters and the rest of the groups (blue and red) have mostly had one or two.  This implies that the significant effects of GMCs mostly arise from strong encounters rather than the cumulative effect of many weak encounters which is in agreement with the conclusions by \citet{Gieles2006}.  Clusters with $\Delta N_p < -0.2$ have typically experienced one or more GMC encounters; the distribution of $N_{\rm enc}$ for this group is significantly broader. The majority of clusters are destroyed by the end of their simulation. Only $\sim 7$ per cent of destroyed clusters experience no encounters, with the distribution of $N_{\rm enc}$ having a long tail. While the larger number of GMC encounters helps drive cluster dissolution, Galactic orbits that undergo more strong GMC encounters also experience stronger heating from the larger-scale tidal field produced by the spiral arms, where the majority of GMCs are located. In total, 13264 clusters are destroyed in the C$_{\rm{R}}$ population, compared to 10789 in the C$_{\rm{N}}$ population.

\subsection{Survivability}
\begin{figure}
 \includegraphics[width=\columnwidth]{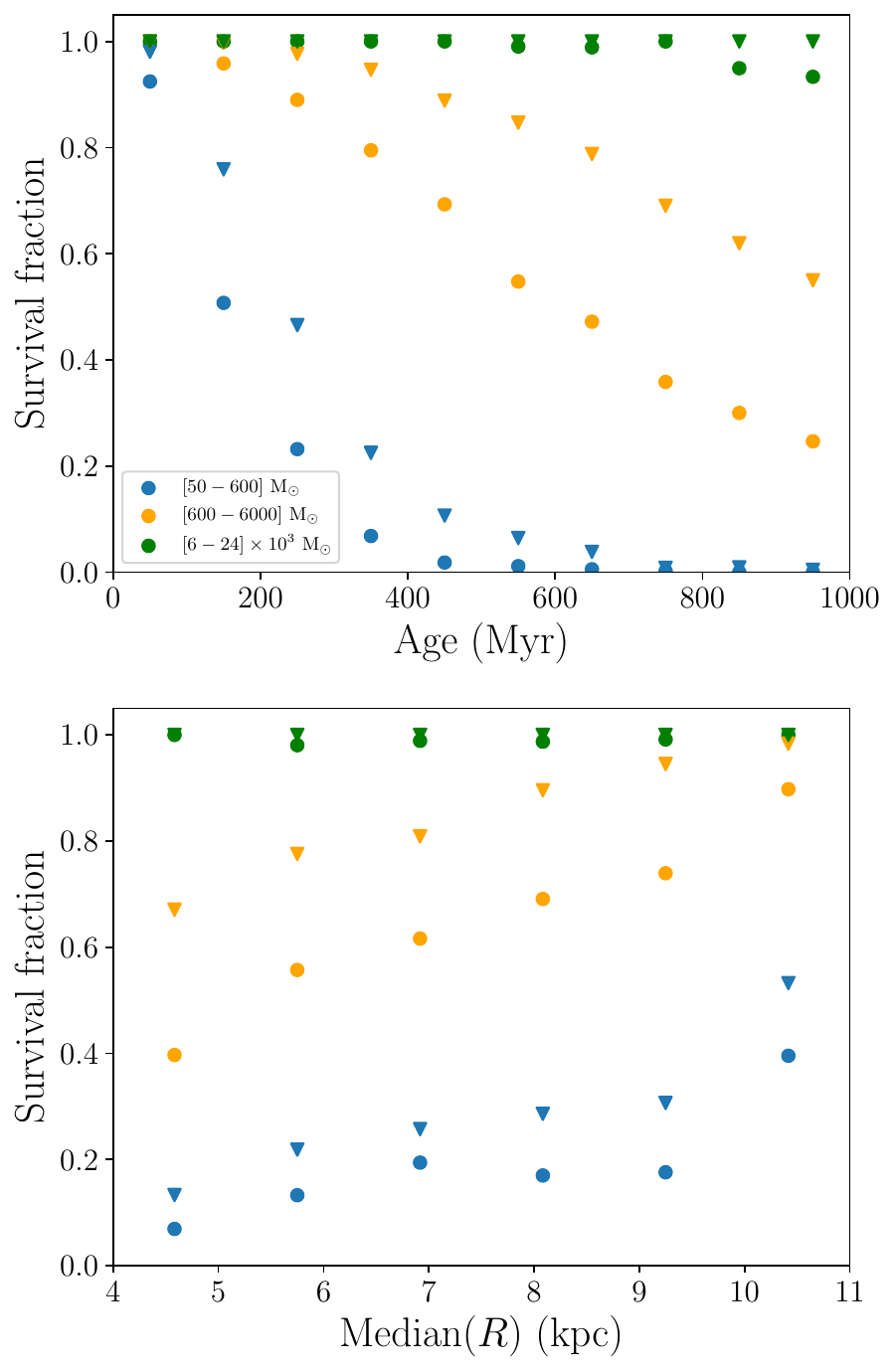}
\caption{The fraction of model clusters that survive as functions of age ({\it top}) and median Galactocentric radius $R$ ({\it bottom}). The clusters are split into three bins by initial cluster mass: low mass ($50<M_{\rm i}/\MO<600$, blue points, lowest survival fraction); intermediate mass ($600<M_{\rm i}/\MO<6000$, orange points, intermediate survival fraction); and high mass ($M_{\rm i}>6000\,\MO$, green points, highest survival fraction). The $\rm C_R$ and $\rm C_N$ clusters are shown as circles and triangles, respectively.}
\label{fig:SF}
\end{figure}

We investigated the survivability of our model clusters by dividing the populations into three groups by initial mass, namely: low-mass ($50<M_{\rm i}/\MO<600)$; intermediate-mass ($600<M_{\rm i}/\MO<6000)$; and high-mass ($M_{\rm i}>6000\,\MO)$. In the top panel of Fig~\ref{fig:SF} we plot the fraction of clusters in each group that survive as a function of age for the C$_{\rm{R}}$ (circles) and C$_{\rm{N}}$ (triangles) populations. For the high-mass clusters, the survival fraction is only affected by GMCs for clusters older than 800\,Myr. For the intermediate mass group the difference starts at $\sim250$\,Myr and grows with age up to $\sim 600$ Myr. The survival of low-mass clusters is most affected by the Galactic environment. While there is a significant difference in survivability of up to about 25 percentage points between 100 and 300\,Myr, the majority of clusters in both simulations have been destroyed after $400$\,Myr. 

The bottom panel of Fig.~\ref{fig:SF} shows the survival fraction as a function of median Galactocentric radius for the three mass groups. For the high-mass clusters, most clusters survive and we see no dependence on Galactocentric radius. For intermediate-mass clusters we see a strong dependence on Galactocentric radius, with the survival fraction increasing roughly linearly with $R$ for both C$_{\rm{R}}$ and C$_{\rm{N}}$ clusters. The greater destruction caused by the higher concentration of GMCs in the inner Galaxy appears to scale along with the destruction caused by the larger-scale potential from the arm and bar, such that the difference in fractional survival between the C$_{\rm{R}}$ and C$_{\rm{N}}$ clusters is roughly constant with $R$. The low-mass clusters also show a strong dependence on Galactocentric radius, with the survival fraction being small other than at $R>10\,{\rm kpc}$. The difference between the C$_{\rm{R}}$ and C$_{\rm{N}}$ populations is generally greater for the intermediate-mass clusters compared to the low-mass clusters. This is because the majority of low-mass clusters dissolve in a few hundred Myr whether GMCs are present or not.

In Section \ref{sec:HuntMethod} we discuss our choice of adopting the Hunt method when defining our simulated clusters in order to make a better comparison with the observations. However, this choice has an impact on the survival fraction and half-mass radii of our simulated clusters.  Using the Hunt method, we find that 7428 clusters survive in our sample, compared to 7815 clusters if we include the stellar energies; the corresponding overall survival fractions are $35.9$ and $37.8$ per cent for the C$_{\rm{R}}$ clusters. The change in the number of clusters that survive for the C$_{\rm{R}}$ and C$_{\rm{N}}$ cluster populations, depending on which cluster definition we use, is roughly $\sim 400$ which means that the choice of method does not impact the conclusions of the effect of GMCs on the cluster population. Including the stellar energies in our cluster definition will mean that stars beyond the tidal radius can still be considered bound and this increases the mean half-mass radius of our sample to $r_h = 3.3\pm 1.3$\,pc from $r_h = 3.1\pm 1.2$\,pc, which is not a significant change.

\begin{figure}
 \includegraphics[width=\columnwidth]{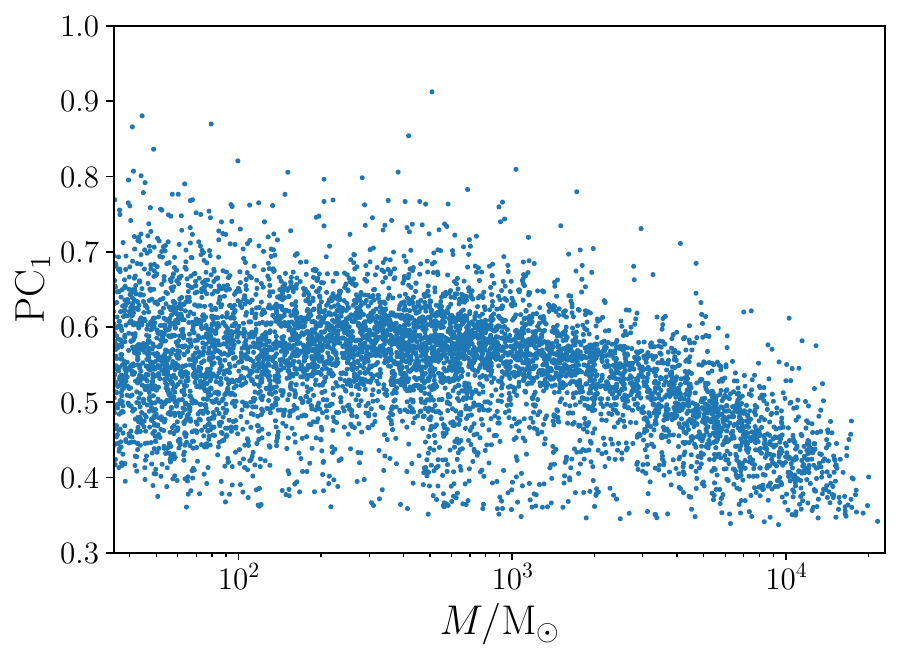}
 \caption{$\mathrm{PC_1}$ as a function of cluster mass for the $\rm C_R$ cluster population. High-mass clusters are generally more spherical than the low-mass clusters.}
 \label{fig:PC1_mass}
\end{figure}

\begin{figure*}
 \includegraphics[width=\textwidth]{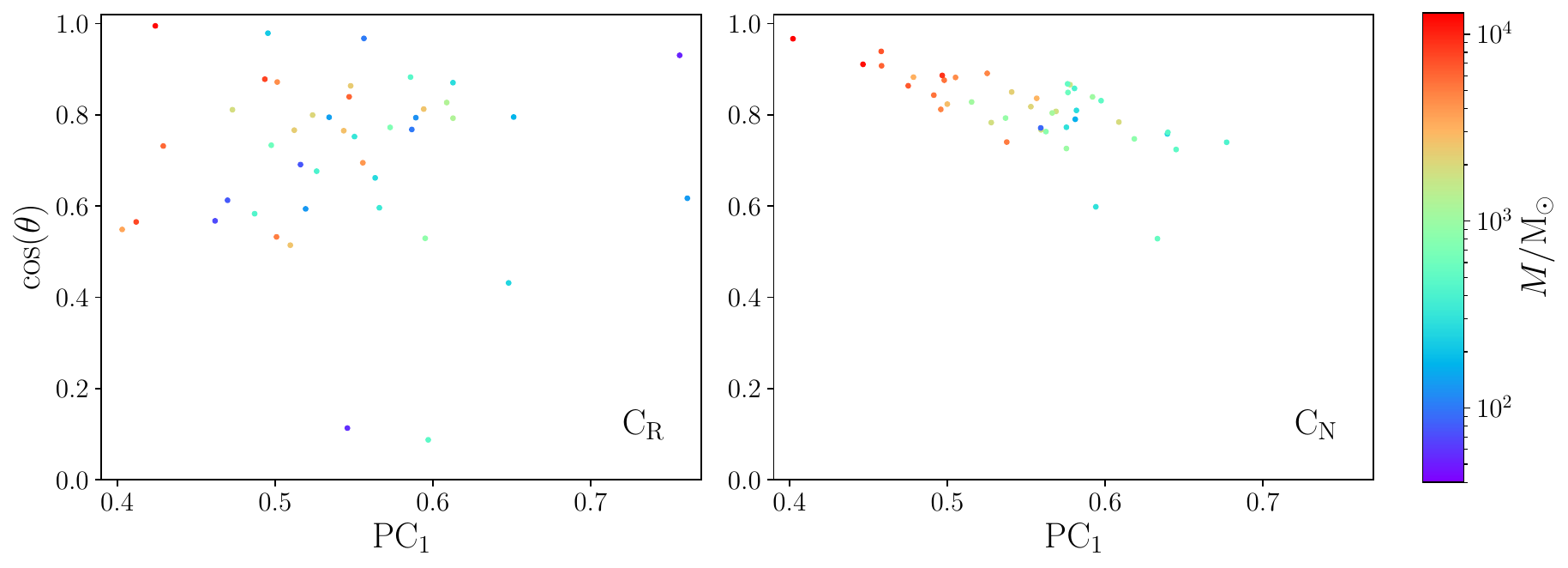}
 \caption{The angle between the major axis and the Galactic centre, $\theta$, for 43 $\rm{C_R}$ clusters (\textit{left}) which have experienced a GMC encounter within the last 20 Myr shown as a function of $\rm{PC_1}$. The clusters are coloured according to their current mass and the corresponding 43 $\rm{C_N}$ clusters are show to the \textit{right}.}
 \label{fig:PCA_cos}
\end{figure*}

\subsection{Cluster morphology}
We can estimate the shape of a cluster by using a Principal Component Analysis (PCA) to analyse the spatial distribution of the stars of our clusters. The PCA method identifies the principal axes where the distribution of stars varies the most, by calculating the eigenvalues and principal components, i.e. eigenvectors, of the covariance matrix of the stellar positions. The shape of each cluster is assessed by calculating the variance
ratios of the eigenvalues of the principal components which for the major axis is given by
\begin{equation}
\mathrm{PC_1} = \frac{\lambda_1}{\sum_{i = 1}^{N_D} \lambda_i}.
\end{equation}
Here $\lambda_i$ is the eigenvalue of the i-th principal component, and $N_D$ is the number of spatial dimensions. If a cluster is close to spherical, $\mathrm{PC_1}$ will be around $1/3$. If instead the cluster is highly elongated, PC$_1$ will be close to 1. We adopt a covariance matrix that includes all stars within $2r_t$ of each cluster, which ensures that we capture the beginning of their tidal tails. Fig. \ref{fig:PC1_mass} shows PC$_1$ of the C$_{\rm{R}}$ clusters as a function of their mass. High-mass clusters are much more spherical since they are less affected by the tidal forces from the Galaxy. Lower mass clusters are more elongated and the spread in PC$_1$ increases which is likely due to stochastic effects in our Milky Way potential, such as the GMCs and the spiral arms. \citet{Vazquez2024} used PCA on a sample of clusters from the open cluster catalogue of \citet{Hunt2023} and found that it is only reliable for clusters within a distance of $\sim 220$ pc due to observational uncertainties of \textit{Gaia}. It is therefore not currently possible to reliably use PCA to compare model clusters to the majority of cluster observations.

Clusters that are steadily losing stars to the Galactic field will generally be elongated towards the Galactic centre, since the stars escape the cluster through the Lagrange points, $L_1$ and $L_2$. If a cluster has had a recent GMC encounter, we would expect that the shape of the cluster to be somewhat elongated, but instead of pointing towards the direction of the Galactic centre, it will have a random orientation. To investigate how big this effect might be, we analysed PC$_1$ for clusters that have had a strong GMC encounter within the last 20 Myr. We also require the cluster mass to be at least $40$ \MO, since this is the minimum mass required to be classified as a cluster in the \citetalias{Hunt2024} catalogue.
This leaves 43 unique clusters that survive in both the C$_{\rm{R}}$ and C$_{\rm{N}}$ simulations. For the 43 clusters, we calculated the angle $\theta$ between their direction towards the Galactic centre and their major axis. Fig. \ref{fig:PCA_cos} shows $\rm{cos}(\theta)$ as a function of PC$_1$ for the C$_{\rm{R}}$ and C$_{\rm{N}}$ clusters which are shown to the left and right, respectively. For the C$_{\rm{N}}$ clusters, the distribution is ordered and we see a clear relation between the shape of the clusters and their orientation towards the Galactic centre. High-mass clusters are more spherical with a clear orientation that points towards the Galactic centre. As the mass of the clusters becomes lower, $\mathrm{cos}(\theta)$ decreases since the stars that escape make up a larger fraction of the total number of stars in the clusters, and therefore have a larger impact on the shape and thus the orientation we derive for the clusters.  The C$_{\rm{R}}$ clusters have a somewhat similar range in their PC$_1$ values, but their orientations in relation to the Galactic centre are much more stochastic. The $\mathrm{PC_1}$ for the high-mass clusters are much more scattered. Some of them remain nearly spherical but the orientation of their tidal tails are no longer pointed towards the Galactic centre. There is also no longer a clear dependence between the cluster mass and $\mathrm{cos}(\theta)$, and some of the low-mass clusters show an orientation which is nearly parallel with their direction towards the Galactic centre. Two of the low-mass C$_{\rm{R}}$ clusters have retained a similar $\mathrm{cos}(\theta)$ but are now very elongated.     

Our analysis shows that it could, in theory, be possible to identify open clusters that have recently had GMC encounters by calculating their PC$_1$ and the orientation of their tidal tails. \citet{Ratzenbock2025} have found evidence for a current ongoing interaction between the stellar disc stream Theia 368, which is currently embedded in the Scorpius-Centaurus association, and the primordial gas mass in the association. Theia 368 shows evidence of it currently experiencing disruption due to this interaction, where stars of Theia 368 that are located further inside Scorpius-Centaurus show higher relative motions which have been altered in the direction towards the association.

\begin{figure*}
    \centering
    \includegraphics[width=0.8\linewidth]{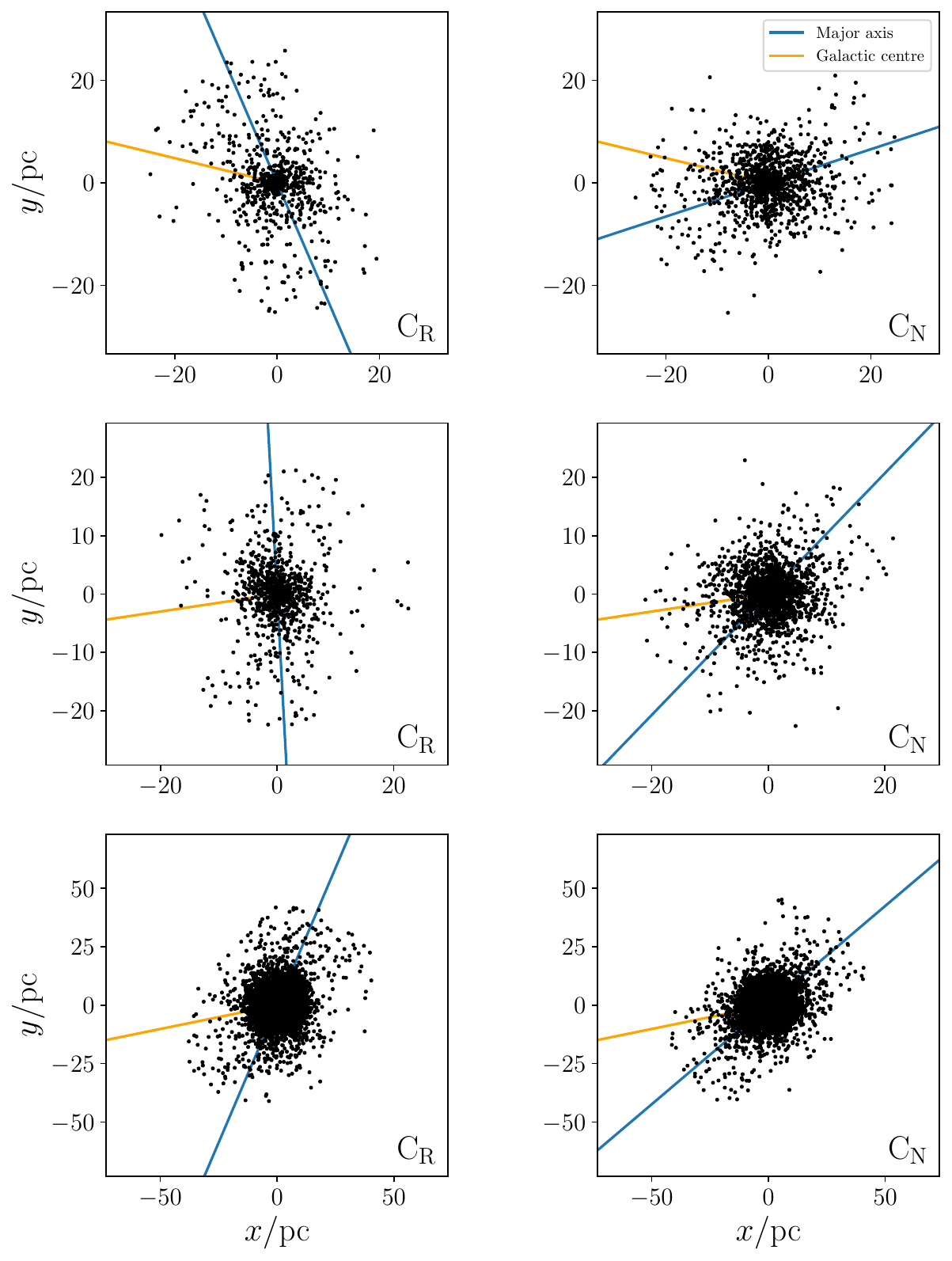}
    \caption{The distribution of stars projected into the $x-y$ plane of the Galaxy for three of the 43 clusters which have had recent GMC encounters. The $\rm{C_R}$ and corresponding $\rm{C_N}$ clusters are shown to the left and right, respectively. The major axis and direction towards the Galactic centre are represented by the blue and orange lines, respectively. The clusters are orbiting clockwise around the Galaxy.}
    \label{fig:XY_3}
\end{figure*}

Fig. \ref{fig:XY_3} shows the distribution of stars projected into the $x-y$ plane of the Galaxy for three clusters which have had recent GMC encounters. The stars of the $\rm{C_R}$ cluster and corresponding $\rm{C_N}$ cluster are shown to the left and right in each row, respectively. The direction of the major axis is indicated by the blue line and the orange line points towards the Galactic centre. The orbital rotations of the clusters in the Galaxy are clockwise. Each of the $\rm{C_N}$ clusters have stars that escape through $L_1$ and $L_2$, which overtakes and lags behind the clusters as a result of the differential rotation of the Galaxy. For the $\rm{C_R}$ clusters, the direction of the tidal tails have been affected by their recent GMC encounters and are less correlated with the direction of the Galactic centre.

\section{Conclusions}
\label{sec:conclusions}
We have investigated the evolution of the open cluster population of the Milky Way, by combining a model of the Milky Way with $N$-body simulations. This has been done for 20692 unique clusters in the mass range $[50-24000]$ \Mo of clusters ages evenly distributed up to an age of 1 Gyr which we compare with the cluster catalogue of \citetalias{Hunt2024}. To investigate the impact that GMC encounters have on the clusters, we performed two $N$-body simulations for each unique cluster: one version where tidal forces produced by the GMCs are present, called C$_{\rm{R}}$ clusters, and another version where the tidal forces of the GMCs are ignored, called C$_{\rm{N}}$ clusters. 

Our cluster masses are based on an initial cluster mass function (ICMF) with a power-law slope of $-2$. When comparing the mass function of our simulated clusters to observations, we find large discrepancies. Instead, if we assume an ICMF described by a Schechter function with a truncation mass of $\sim 9000$ \MO, we get much better agreement with the observed population. We therefore adopt the truncated cluster populations, which we refer to as C$_{\rm{RT}}$ and C$_{\rm{NT}}$ in the paper. 

We explore the evolution of the mass function as a function of which truncation mass is applied to the ICMF. For truncation masses above $\sim 15\,000$ \MO, the slopes of the mass function do not change, since the distribution of clusters is less affected by the high-mass clusters, since they contribute very little to the mass function in terms of the actual number of clusters. Below a truncation mass of $\sim 15\,000$ \MO, we see that the slope of the mass function gets steeper, except for young clusters which have not had time to evolve and therefore have no change in their initial mass slope of $-2$. However, for a truncation mass below $\sim 8\,000$ \MO, clusters in the age range between 100 and 200 Myr have a mass function slope significantly steeper than $-2$ which we attribute to stochastic effects since our sample size becomes too low. If a truncation mass is not applied to the ICMF, the evolution of the slope of the mass functions flattens over a time-scale which is twice as short of $(1.12\pm 0.27)$ Gyr, compared to the C$_{\rm{RT}}$ clusters where the time-scale is $(2.11\pm 1.49)$ Gyr with the observed time-scale being $(2.10\pm 0.53)$ Gyr.   

We find that the mass function for the total population of clusters with ages less than 1 Gyr is well described by either a broken power-law or a Schechter function, but we find the best agreement between observations and our C$_{\rm{RT}}$ population with the Schechter function fit. We find a slope of $\beta_s = -1.54\pm 0.03$ and truncation mass of $M_t = (4.04\pm 0.49) \times 10^3$ \MO, compared to the observed values of $\beta_s = -1.50\pm 0.04$ and $M_t = (3.52\pm 0.47) \times 10^3$ \MO. A non-truncated ICMF results in a similar slope, but matches less well at higher masses.

The inferred truncation mass, $M_t$, in the Milky Way is very low compared to other spiral galaxies \citep{Adamo2015,Hollyhead2016,Adamo2017, Johnson2017,Messa2018}. Environmental investigations performed on some of these galaxies show that $M_t$ can decrease significantly from the inner to the outer regions the galaxies \citep{Adamo2015, Adamo2017, Messa2018b} and it has been suggested by \citet{Johnson2017} that $M_t$ could be related to the star formation rate surface density. The observed clusters in the Milky Way are mostly within 3\,kpc of the Sun: \citetalias{Hunt2024} estimate that their catalogue contains $\sim 4$ per cent of the total cluster population of the Milky Way. We therefore suggest that the low $M_t$ observed in the Milky Way might be a poor description of the total open cluster population of the Milky Way.

The age functions for all cluster populations are best described by broken power-laws with slopes $\gamma_1$, $\gamma_2$, and a break point $t_{\mathrm{break}}$. The age function of the C$_{\rm{RT}}$ population suggests that there is no significant cluster disruption in the first $\sim 50$ Myr with $\gamma_1 = -0.07\pm 0.06$, whereas the observations indicate that cluster disruption occurs within $\sim 1$ Myr, with a corresponding $\gamma_1 = -0.58\pm 0.05$. The observed age function also shows a greater disruption of clusters older than $\sim200$ Myr with $\gamma_2 = -2.16\pm 0.22$, compared to our simulations with $\gamma_2 = -1.25\pm 0.05$. 

To investigate if these discrepancies between our simulations and the observations might be caused by incompleteness of the low-mass clusters, we analysed the age function for clusters in the mass range $[600,6000]$ \MO. We find that this causes the breakpoint for the broken power-law fit to match between our cluster models and the observations. However, there is still a substantial discrepancy in the slopes of the age functions. The age uncertainties in \citetalias{Hunt2024} are fairly high for the older clusters, but cannot explain the discrepancy between the observed age function and our simulations. The early disruption of clusters seen in the observations suggests that they might be born in a high virial state, while all clusters in our simulations are born in virial equilibrium. We test this hypothesis by re-running $N$-body simulations of the surviving C$_{\rm{RT}}$ clusters with final masses above 50 \Mo and ages younger than 100 Myr. This is done for two different initial virial ratios of 0.75 and 1.25. We find that clusters born with a initial virial parameter of 1.25 match the observations best when their age function is fitted to a broken power-law with resulting values of $\gamma_1 = -0.49\pm 0.09$ and $\gamma_2 = -2.15\pm 0.22$. These values match the observations well, however, the breakpoint does not align with the observations, and this might suggest that the initial virial state of the clusters are correlated with their initial mass.

Out of the 20692 simulated clusters, we find that 7428 and 9903 survive for the C$_{\rm{R}}$ and C$_{\rm{N}}$ clusters, respectively. By directly comparing each of the clusters in the C$_{\rm{R}}$ and C$_{\rm{N}}$ populations, we find that clusters with the largest difference in their number of stars are low-mass ($<600$ \MO) clusters that have undergone one or more GMC encounters. By comparing the survival fraction for the C$_{\rm{R}}$ and C$_{\rm{N}}$ clusters as a function of age and median Galactocentric radius, we find that GMCs have the greatest effect on the low-mass cluster survivability in the first $\sim 200$ Myr of their lifetime and that all low-mass clusters are destroyed on a time-scale of $\sim 500$ Myr. Low-mass clusters which spend most of their time at Galactocentric radii of $\sim 4 - 5$ kpc show little difference in survivability between the C$_{\rm{R}}$ and C$_{\rm{N}}$ clusters, which indicates that the GMCs have little affect on the cluster destruction. This is because these clusters would be destroyed by the rest of the Galactic tidal field, whether they would have had GMC encounters or not. Intermediate-mass ($600-6000$ \MO) clusters are most impacted by GMCs, since they survive for longer and hence have a greater chance of undergoing close encounters with GMCs. These clusters also show a linear increase in their survival fraction as their distance to the Galactic centre increases.

We perform a Principal Component Analysis (PCA) on the stars of the C$_{\rm{R}}$ and C$_{\rm{N}}$ clusters and find a clear relation between the shape and mass of clusters. Clusters with high masses are generally more spherical, whereas low-mass clusters are more elongated. Clusters that have experienced a strong GMC encounter within the last $20$ Myr have tidal tails which have a random orientation in relation to the direction of the Galactic centre, whereas their C$_{\rm{N}}$ counterparts show a clear relation between their morphology, mass and the direction of their tidal tails.

\section*{Acknowledgements}

We would like to thank Carlos Viscasillas Vázquez, Laura Magrini, and Emily Hunt for valuable discussions and suggestions that helped improve this paper. We would also like to thank the anonymous
referee for input that has helped strengthen this paper. The $N$-body computations were enabled by resources provided by LUNARC, The Centre for Scientific and Technical Computing at Lund University, which were possible thanks to grants from The Royal Physiographic Society of Lund.

\section*{Data Availability}
The data underlying this article will be shared on reasonable request
to the corresponding author.


\bibliographystyle{mnras}
\bibliography{references} 



\bsp	
\label{lastpage}
\end{document}